\documentclass[pre, twocolumn, superscriptaddress,showpacs]{revtex4-1}

\usepackage{fullpage}
\usepackage{amsmath, amsthm, amssymb}
\usepackage{graphicx}
\usepackage{color}
\usepackage{bm}
\usepackage{hyperref, verbatim}

\begin{document}

\title{Transport properties of topological superconductor-Luttinger liquid junctions: a real-time Keldysh approach}
\author{Roman M. Lutchyn}
\affiliation{Station Q, Microsoft Research, Santa Barbara, CA 93106-6105}
\author{Jacob H. Skrabacz}
\affiliation{Department of Physics, University of California, Santa Barbara, CA, 93106}

\date{\today}

\begin{abstract}
Inspired by a recent experimental observation of the zero-bias tunneling conductance in superconductor-semiconductor nanowire devices~\cite{Mourik2012},  we consider here transport properties of the junctions consisting of a nanowire (Luttinger liquid) coupled to a topological superconductor characterized by the presence of Majorana zero-energy end states. The presence of the Majorana modes leads to a quantization of the zero-bias tunneling conductance at zero temperature. In order to understand this phenomenon, we have developed a framework, based on real-time Keldysh technique, which allows one to compute tunneling conductance at finite temperature and voltage in a realistic experimental setup. Our approach allows one to understand this transport phenomenon from a more general perspective by including the effect of interactions in the nanowire, which sometimes results in a drastic departure from the non-interacting predictions. Thus, our results provide a key insight for the tunneling experiments aiming at detecting Majorana particles in one-dimensional nanowire devices.
\end{abstract}

\maketitle

\section{Introduction}

The search for topological phases in nature has been an active and exciting pursuit. Recent experimental progress~\cite{Mourik2012, Rokhinson2012, Das2012, Deng2012, Fink2012} aimed at detecting Majorana zero-energy modes (Majoranas) in one-dimensional superconductor/semiconductor nanowire devices has excited the whole physics community~\cite{Reich, Brouwer_Science, Wilczek2012}. The reason for this excitement is two-fold. First of all, Majorana modes are predicted to obey non-Abelian braiding statistics~\cite{Moore1991, Nayak1996, ReadGreen, ivanov, BondersonNonAbelianStatistics} and, thus, their discovery is of fundamental importance for all of physics. Second, this property of Majoranas (or Ising anyons) is at the heart of topological quantum computing schemes~\cite{kitaev, TQCreview}. Indeed, the Majorana modes localized at the opposite ends of the nanowire provide a one-dimensional topologically protected two-level system~\cite{Kitaev:2001}. The network of Majorana wires can be used for topological quantum information processing~\cite{AliceaBraiding, SauWireNetwork, ClarkeBraiding, TopologicalQuantumBus}. The topological approach to quantum
computing is very promising because it provides a way to protect information from local errors by encoding it in non-local degrees of freedom of topologically ordered systems~\cite{kitaev, TQCreview}.

It has been predicted that Majorana zero-energy modes can emerge quite naturally in spinless p-wave superconductors~\cite{ReadGreen, Kitaev:2001}. Experimental realization of a spinless p-wave superconducting state, which is robust against various perturbations inherently present in any solid-state system (e.g. disorder),  is a non-trivial task. One attractive class of experimental proposals involves a conductor with strong spin-orbit coupling allowing one to enslave the spin degree of freedom and an s-wave superconductor~\cite{Fu:2008, MajoranaQSHedge, Sau, Alicea, 1DwiresLutchyn, 1DwiresOreg}. The experimental search of Majoranas in solid-state systems has been focused so far on one-dimensional superconductor/semiconductor nanowire devices~\cite{1DwiresLutchyn, 1DwiresOreg} because of the simplicity and robustness of these systems. (We refer a reader to the following review articles~\cite{BeenakkerReview, AliceaReview} for a detailed discussion and comparison of different experimental proposals.) The presence or absence of Majorana zero-energy modes in one-dimensional topological superconductors can be detected by measuring tunneling local density of states at the ends of the wire. Shortly after the Majorana wire proposals~\cite{1DwiresLutchyn, 1DwiresOreg}, there have been several experimental reports~\cite{Mourik2012, Das2012, Deng2012, Fink2012} of zero-bias tunneling conductance in superconductor/semiconductor nanowire devices, which can be interpreted as a signature of Majorana zero-energy modes. This interpretation of the observed zero-bias tunneling conductance relies on theoretical predictions based on non-interacting models for one-dimensional semiconductor-superconductor heterostructures ~\cite{ZeroBiasAnomaly0,ZeroBiasAnomaly1,ZeroBiasAnomaly2,ZeroBiasAnomaly3,ZeroBiasAnomaly31, ZeroBiasAnomaly4,ZeroBiasAnomaly5,ZeroBiasAnomaly6,ZeroBiasAnomaly61, ZeroBiasAnomaly7}.

\begin{figure}[tbp]
\centerline{\includegraphics[width=3.3in]{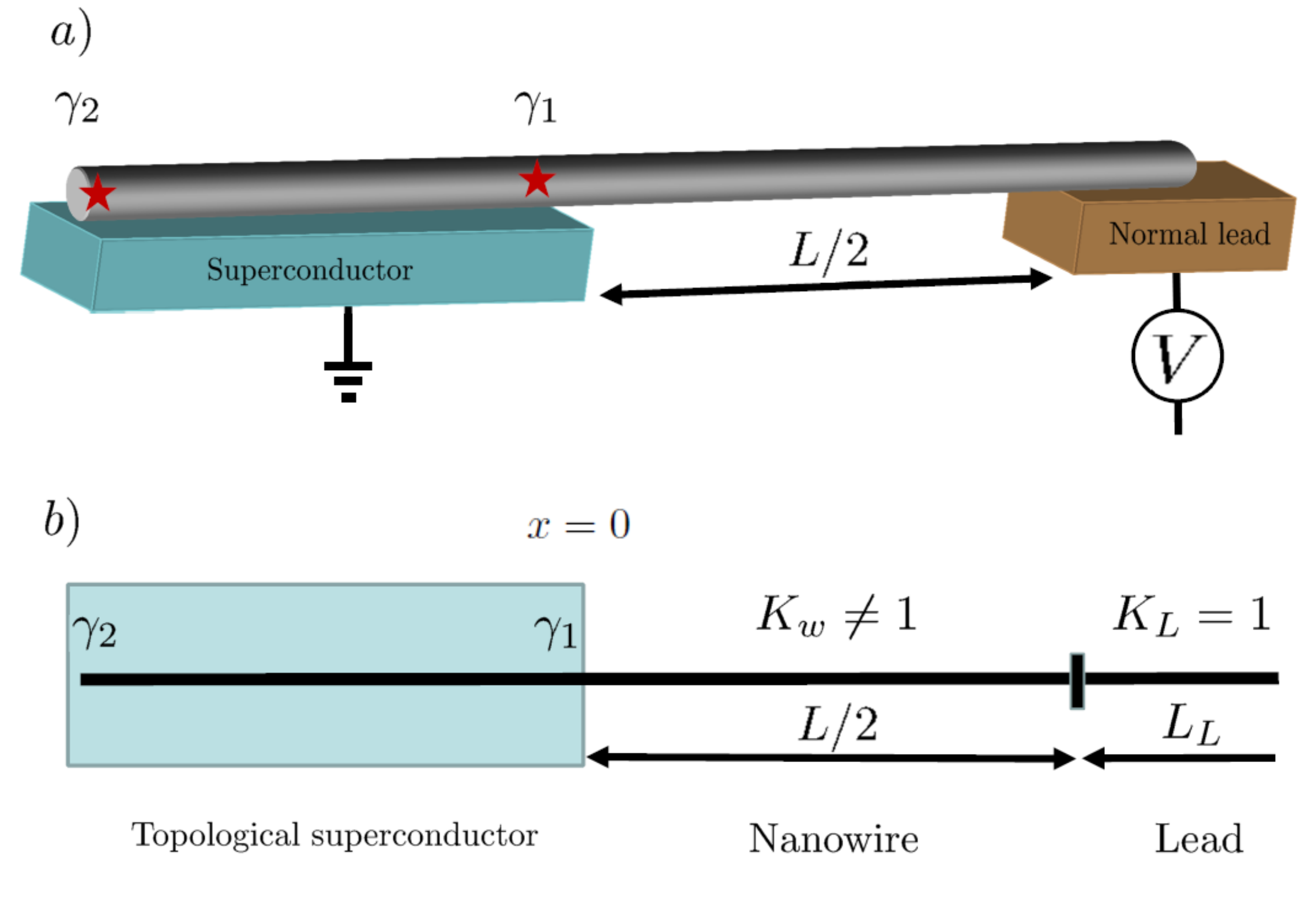}}
\caption{a) Schematic picture of the semiconductor/superconductor hybrid device in Delft experiment~\cite{Mourik2012}. b) Schematic picture of the theoretical model for the corresponding device. }
\label{Fig:setup}
\end{figure}

In this paper we revisit transport properties of a TP SC/LL junction from a much more general perspective using a combination of renormalization group (RG) and real-time Keldysh techniques. Our approach allows one to take into account electron-electron interactions, which are particularly important in one-dimensional systems~\cite{Giamarchi:2004}, and compute transport properties of the superconductor-semiconductor nanowire devices as a function of various parameters such as temperature, voltage, and length of the nanowire. The system of interest involves a topological superconductor coupled to a Luttinger liquid representing the uncovered portion of a nanowire, see Fig.~\ref{Fig:setup}a for the schematic plot of the experimental setup~\cite{Mourik2012}. Thus, the simplest theoretical model for the tunneling experiment~\cite{Mourik2012} consists of a topological superconductor coupled to a one-dimensional nanowire of length $L/2$. The other end of the nanowire is connected to a semi-infinite normal-metal lead which is a part of the circuit for measuring tunneling conductance. Here the metallic lead is described as a three-dimensional Fermi liquid where interaction effects between quasiparticles can be neglected. For our purposes, it can be represented as a non-interacting Luttinger liquid with $K_L=1$. Therefore, in order to study transport properties of such a system, we introduce an inhomogeneous LL interaction parameter $K(x)$ which adiabatically interpolates between $K(x)=K_w$ and $K(x)=K_L$ in the nanowire and leads, respectively. The schematic plot of the experimental setup is shown in Fig.~\ref{Fig:setup}b.

In the absence of the lead, the properties of TP SC/LL junctions have been studied in Ref.~\cite{Fidkowski2012} where it was shown that the quantization of the tunneling conductance in units of $2e^2/h$ is robust even in the presence of electron-electron interactions in the wire. In particular, for weaker electron-electron interactions ($K_w>1/2$), the system flows to the so-called {\it perfect Andreev reflection} fixed point where the zero-bias tunneling conductance is equal to twice the conductance quantum at zero temperature. We note here that in the case of a non-topological superconductor coupled to a Luttinger liquid~\cite{maslov, Affleck2000, LDOSdivergence}, the {\it perfect Andreev reflection} fixed point is unstable for any repulsive interparticle interactions in the wire, i.e. for $K_w < 1$. Thus, there is a qualitative difference between TP SC- and NTP SC-LL junctions.

Building on the previous results~\cite{Fidkowski2012}, we study here the transport properties of a TP SC/LL junction in the presence of the leads. Specifically, we derive the Keldysh action for the boundary theory describing such a junction and then compute its transport properties. We first show unambiguously that tunneling conductance is quantized in units of $2e^2/h$, as expected from the previous studies~\cite{Apel1982, KaneFisher1, Wen1991, Matveev1993, Furusaki1993, Safi_Schulz, Maslov1995,  Ponomarenko1995, Oreg1995, Alekseev1996, Chamon1996, DolciniPRL'03, DolciniPRB'05, Thomale}. Next, we compute temperature- and voltage-dependent corrections to the tunneling conductance and discuss their scaling in the presence of the leads. We show that in some instances the presence of interactions in the nanowire results in a qualitatively different dependence of the tunneling conductance on physical parameters compared to the non-interacting predictions. Thus, our results provide important information for the Majorana tunneling experiments, particularly since the maximum zero-bias peak value is smaller than $2e^2/h$ by an order of magnitude in current tunneling experiments\cite{Mourik2012, Das2012, Deng2012, Fink2012}. The nature of the reduced tunneling conductance from the quantized value $2e^2/h$ at finite temperature and voltage is an open question right now.

The theoretical framework developed in this paper has broader impacts and can be applied to other quantum impurity problems, see, {\it e.g. }, lecture notes~\cite{Affleck2008} and references therein. We note that the presence of the leads breaks integrability of the boundary sine-Gordon model, and the exact solution~\cite{Fendley1995} is not available in this case. Thus, perturbative calculations such as the one developed here are particularly useful. Furthermore, the real-time Keldysh approach is suitable for studies of time-dependent phenomena. Indeed, it should be straightforward to compute noise correlation functions using this method.

The paper is organized as follows. In Sec.~\ref{sec:qual}, we briefly review previous results for the tunneling conductance in one-dimensional Majorana wires. In Sec.~\ref{sec:Keldysh}, we introduce Keldysh formalism and compute transport properties of a TP SC/LL junction first for an infinite LL and then for a finite LL coupled to a metallic lead as shown on Fig.~\ref{Fig:setup}b. In Sec.~\ref{sec:mainresults}, we summarize our main findings and provide a qualitative explanation of the temperature and voltage dependence of the conductance using a two-step RG procedure. The technical details are relegated to the Appendices.

\section{Discussion of the previous results}\label{sec:qual}

\subsection{Zero-bias anomaly and tunneling conductance in the non-interacting limit}

In this section, we briefly review previous results for transport calculations in the non-interacting model of TP SC/LL junctions \cite{ZeroBiasAnomaly0,ZeroBiasAnomaly1,ZeroBiasAnomaly2,ZeroBiasAnomaly3,ZeroBiasAnomaly31, ZeroBiasAnomaly4,ZeroBiasAnomaly5,ZeroBiasAnomaly6,ZeroBiasAnomaly61, ZeroBiasAnomaly7}. The presence of the lead is not important here since $K_w=K_L=1$, and we assume that $L\rightarrow \infty$.

Without loss of generality, we focus here on the transport properties of a spinless nanowire coupled to a spinless 1D superconductor. For simplicity, we consider below a single-band nanowire but our results should also be applicable to multi-band superconductor/semiconductor heterostructures as well~\cite{WimmerMultichannel, 1DwiresPotter, 1DwiresLutchyn2}. Throughout this paper, we will assume that the superconductor is fully gapped and sufficiently long so that the splitting between Majorana end states can be neglected. Therefore, the LL is effectively coupled only to $\gamma_1$, see Fig.\ref{Fig:setup}, which is the only relevant low-energy superconducting degree of freedom in the problem. Taking into account the aforementioned assumptions, the effective model for the topological superconductor-LL junction becomes ($\hbar=1$)
\begin{align}
H &= H_0 + \delta H \label{eq:H0},\\
H_0& = \int_0^\infty dx\left(-i v \psi_R^\dagger \partial_x \psi_R + i v \psi_L^\dagger \partial_x \psi_L\right), \label{eq:H0}
\end{align}
where $v$ is the Fermi velocity and $\psi_{R/L}^\dagger$ represents right/left-moving excitations near the Fermi energy.  It is convenient to perform an unfolding procedure and rewrite $H_0$ in terms of a single fermion field $\psi(x)$ defined over all $x$ as follows:
\begin{eqnarray}
  \psi(x) = \left\{ \begin{array}{rl}
 \psi_R(x), & x > 0 \\
 \psi_L(-x), & x < 0.
       \end{array} \right.
   \label{eq:psi}
\end{eqnarray}
In terms of $\psi(x)$, $H_0$ becomes simply
\begin{equation}
  H_0 = \int_{-\infty}^\infty dx\left(-i v \psi^\dagger \partial_x \psi \right).
  \label{eq:H0psi}
\end{equation}
The Hamiltonian $\delta H$ in Eq.\eqref{eq:H0} represents boundary terms. As shown in Ref.~\cite{Fidkowski2012}, the most important processes from the RG perspective are tunneling to Majorana mode and electron backscattering from the boundary defined by the amplitudes $\lambda_M$ and $\lambda_N$, respectively. The boundary Hamiltonian is given by
\begin{align}
\!\delta H \!=\! \int _{-\infty}^\infty\! dx \! \left(\frac{\lambda_M}{\sqrt{2}} \gamma_1(\psi^\dagger-\psi) + 2\lambda_N\psi^\dagger \psi\right)\!\!\delta(x).
\end{align}

The tunneling current for this setup can be calculated using Landauer formalism. The subgap transport at low temperature and voltage $T, eV \ll \Delta$ is determined by the Andreev reflection probability $T(E)$~\cite{BeenakkerReview, AliceaReview}:
\begin{align}\label{eq:Itun}
I=\frac{2e}{h}\int dE (f_{F}(E-eV)-f_F(E))T(E),
\end{align}
where $f_F(E)$ is the Fermi distribution function. A straightforward calculation of the Andreev reflection probability $T(E)$ yields
\begin{align}\label{eq:T}
T(E)=\frac{\Gamma^2}{\Gamma^2 + E^2},
\end{align}
where $\Gamma=v(\lambda_M/v_F)^2/(1 +(\lambda_N/v_F)^2)$, see Ref.~\cite{Fidkowski2012} for details. The differential tunneling conductance $G=dI/dV$ can be calculated in various limits using Eqs.\eqref{eq:Itun} and \eqref{eq:T}. At zero temperature $T=0$, one arrives at
\begin{align}\label{eq:conductance_noni_V}
\frac{G}{2e^2/h}=\left\{
              \begin{array}{c}
                \frac{\Gamma^2}{(eV)^2},  eV \gg \Gamma, \\ \\
                1-\frac{(eV)^2}{\Gamma^2}, eV \ll \Gamma.\\
              \end{array}
\right.
\end{align}
At finite temperature $T\gg eV$, the conductance $G$ is given by
\begin{align}\label{eq:conductance_noni_T}
\frac{G}{2e^2/h}=\left\{
              \begin{array}{c}
                \frac{\pi \Gamma}{4T},  T \gg \Gamma, \\ \\
                1-\frac{\pi^2 T^2}{3\Gamma^2}, T \ll \Gamma.\\
              \end{array}
\right.
\end{align}
One can notice that the scaling of the conductance with temperature and voltage in the limit max$\{eV, T\} \gg \Gamma$ is different, which is rather puzzling since the scaling theory predicts the same power-law dependance for temperature and voltage~\cite{Fidkowski2012}. We will show below how to resolve this contradiction. It turns out that the noninteracting result is non-generic and as soon as we include interactions in the nanowire (i.e. $K_w\neq 1$), the scalings of the conductance with $T$ and $eV$ become the same.

\subsection{The phase diagram of a spinless TP~SC/LL junction}\label{sec:inter}

In this section, we discuss how interparticle interactions in the nanowire affect transport properties of the TP SC/LL junction. Interaction effects in the topological superconductor itself have been previously discussed in Refs.\cite{MajoranaInteractions, Miles, lutchyn_fisher, Sela, Fidkowski2011, LobosPRL'12}. The main conclusion of these publications is that weak interparticle interactions do not destroy the topological superconducting state but rather lead to the renormalization of some physical parameters, such as induced coherence length. Therefore, the simplified description of a Majorana wire in terms of two delocalized zero-energy modes captures the basic physics of an interacting topological superconductor. Consequently, in the remainder of the paper, we focus on the effect of interactions in the LL which can significantly affect transport properties of the TP SC/LL junction. We first briefly review the phase diagram for topological SC/LL junctions obtained in Ref.~\cite{Fidkowski2012}. Throughout this section, we assume for simplicity that the length of the wire $L_w\rightarrow \infty$ in which case one can use the RG approach to obtain the phase diagram. A major virtue of the RG approach is that it allows one to extract universal transport signatures in the presence of interactions by computing the scaling dimension of various perturbations. Thus, one can make rather general statements about transport properties of TP SC/LL junctions.

\begin{figure}[tbp]
\centerline{\includegraphics[width=3.3in]{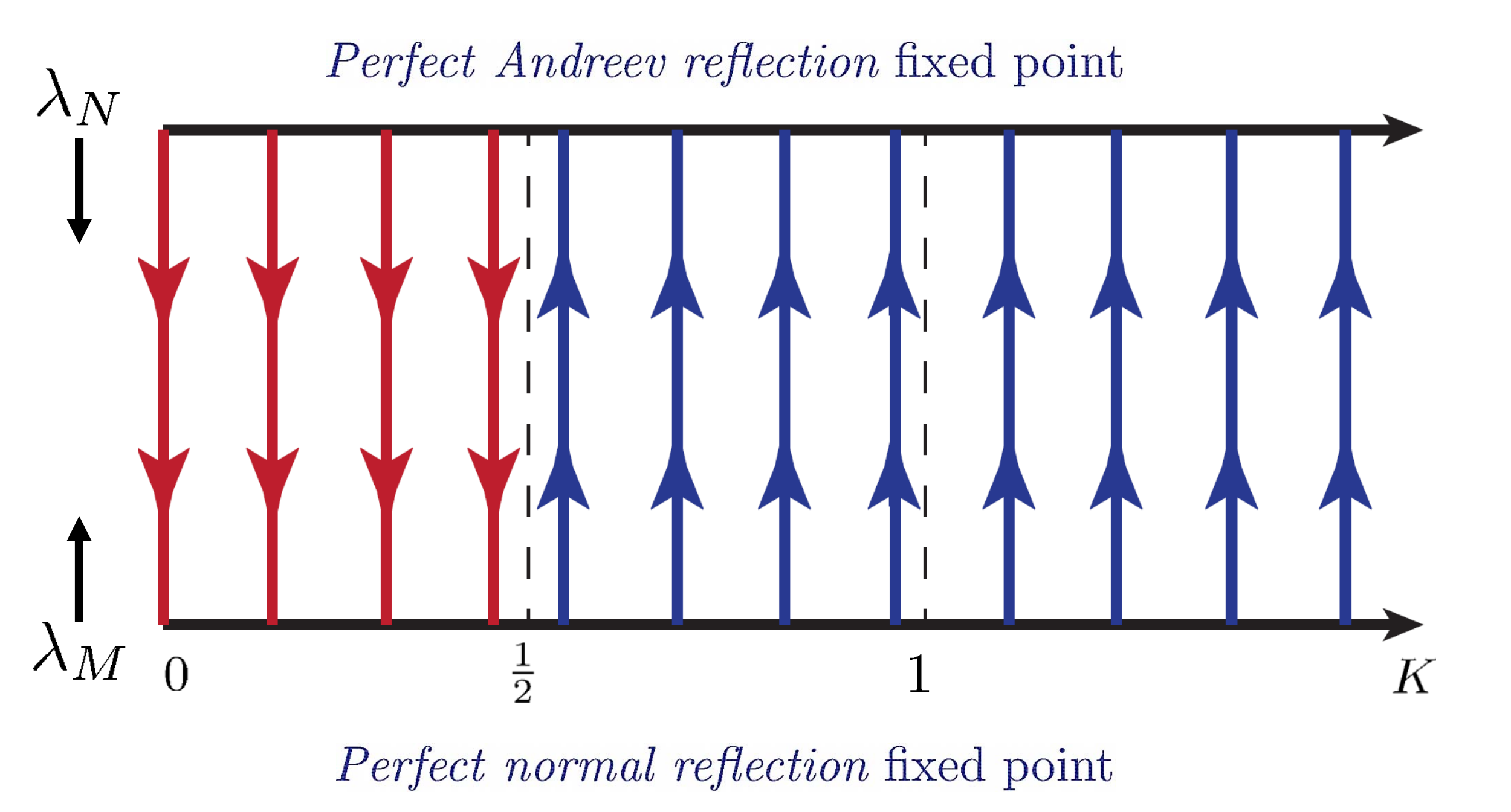}}
\caption{ Phase diagram for TP SC/LL junction as a function of LL liquid parameter in the wire $K$.}
\label{fig:phase}
\end{figure}

It has been shown in Ref.~\cite{Fidkowski2012} that there are two fixed points to which the system renormalizes depending on the LL parameter $K$, see Fig.\ref{fig:phase}.
These two fixed points correspond to {\it perfect normal} and {\it perfect Andreev reflection} and represent two different boundary conditions at the junction (i.e. at $x=0$). The former corresponds to $\psi_R(x=0)=\psi_L(x=0)$ which, according to the bosonization convention $\psi_{R/L}(x)\propto e^{i(\theta(x) \pm \phi(x))}$~\cite{Giamarchi:2004}, results in the pinning of the $\phi$-field at the boundary, i.e. $\phi(x=0)=0{\, \rm mod \, } \pi $. The corresponding imaginary-time boundary action at the normal-reflection fixed point is defined in terms of the dual fluctuating field $\Theta$:
\begin{align}
S_N=\frac{K}{2\pi}\int \frac{d\omega}{2\pi}|\omega||\Theta(\omega)|^2.
\end{align}
Here $K$ is Luttinger liquid parameter characterizing the strength of the interactions and $\Theta(\tau)\equiv \theta(\tau, x=0)$. At the {\it perfect normal reflection} fixed point, the current through the system is zero since electrons are completely backscattered from the junction.

In the opposite limit (i.e. at the {\it perfect Andreev reflection} fixed point), the field $\Theta$ is pinned at the boundary and the corresponding low-energy action is given by
\begin{align}
S_A=\frac{1}{2\pi K}\int \frac{d\omega}{2\pi}|\omega||\Phi(\omega)|^2
\end{align}
where $\Phi(\tau)\equiv\phi(\tau, x=0)$. In this limit, the electric current is finite at zero temperature and is carried by electrons undergoing resonant Andreev reflection.

We now discuss the stability of different fixed points in the presence of interparticle interactions. Following the RG analysis~\cite{Fidkowski2012}, one finds that the {\it perfect Andreev reflection} fixed point is stable for $K>1/2$. In this domain, the leading relevant operator corresponds to the Majorana-coupling boundary term
\begin{align}
  S_M &= 2\lambda_M \int d\tau \sigma^x \cos\Theta.
\label{eq:Majorana}
\end{align}
Here $\sigma_x$ is the Pauli matrix associated with the two-fold ground state degeneracy of the topological superconductor. As shown in Ref.\cite{Fidkowski2012}, $\sigma_x$ is a conserved
quantity, i.e. it has no imaginary time dynamics and for our purposes can be ignored.

To understand the stability of the {\it perfect normal} and {\it perfect Andreev reflection} fixed point, we now analyze the effect of various perturbations. Starting at the normal-reflection fixed point, one finds that the leading relevant perturbation corresponds to tunneling onto the Majorana zero-energy mode. The RG flow of the coupling constant $\lambda_{M}$ reads
 \begin{align}
  \frac{d\lambda_M}{dl}&=\left(1-\frac{1}{2K}\right)\lambda_M.
\label{eq:Majorana}
\end{align}
Thus, Majorana tunneling destabilizes the {\it perfect normal reflection} fixed point for $K>1/2$ and drives the system towards the {perfect Andreev reflection} fixed point. In this domain of the LL parameter $K$, the tunneling conductance approaches the familiar quantized value $G = 2e^2/h$ at zero temperature and zero bias even in the presence of the arbitrary symmetry-allowed boundary couplings.

In the opposite limit $K< 1/2$, the {\it perfect Andreev reflection} fixed point becomes unstable due to the normal reflection at the junction. This process corresponds to the most relevant operator for $K<1/2$. The corresponding imaginary-time action at the boundary reads
\begin{align}
S_R=-\lambda_N \int d\tau \cos(2\Phi).
\end{align}
Indeed, the RG flow for $\lambda_N$ reads
\begin{align}\label{eq:lambda_N}
\frac{d\lambda_N}{dl}=\left(1-2K\right)\lambda_N.
\end{align}
As expected, one finds that the tunneling conductance in this domain scales to zero at zero temperature and voltage. The results concerning the stability of different fixed points are summarized in Fig.\ref{fig:phase}.

\section{Tunneling conductance calculation using real-time Keldysh formalism}\label{sec:Keldysh}

To make a connection with tunneling experiments, it is important to take into account the corrections to the conductance due to finite temperature and voltage, finite length of the nanowire $L/2$, as well as the presence of the leads, see Fig.~\ref{Fig:setup}. Henceforth, we focus on the experimentally relevant regime $K>\frac{1}{2}$ where the {\it perfect Andreev reflection} fixed point is stable. Throughout the paper we will use Keldysh formalism (see Appendix~\ref{App:A} for details and our conventions) which, as we argue below, is particularly suitable for the problem at hand. First of all, it allows one to maintain the analytical structure of the perturbation theory at every step of the calculation and avoids the subtleties with analytical continuation appearing in Matsubara formalism. Second, one can distinguish unambiguously the frequency, temperature, and voltage dependence of the response functions, which enables one to go beyond the scaling theory and obtain the conductance as a function of an arbitrary ratio between voltage and temperature.

In the spirit of the Kane-Fisher approach~\cite{KaneFisher1, KaneFisher2, KaneFisher3}, we first integrate out bulk LL modes and derive a boundary sine-Gordon model. We then assume that the system renormalizes to a particular critical point and study the effect of the leading relevant/irrelevant perturbations in the vicinity of the critical point. This allows one to understand the transport properties as a function of voltage and temperature as well as other physical parameters.

\subsection{Conductance near {\it perfect normal reflection} fixed point}
The relevant model for the TP SC/LL near the unstable {\it normal reflection} fixed point reads
\begin{align}
Z=\int D\Theta & \exp\left[-\frac{K}{2\pi}\int \frac{d\omega}{2\pi} |\omega| |\Theta(\omega)|^2\right.\\
& \left.+\lambda_M \int d\tau \cos\left(\Theta(\tau)+\frac{a(\tau)}{2}\right)\right].\nonumber
\end{align}
Here $a(\tau)$ represents  the gauge field (i.e. $eV(t)=\frac{d a(t)}{dt}$ in real time). The current through the junction is given by $I(\tau)=\delta \ln Z/\delta a(\tau)$\cite{KaneFisher2}.

We proceed with the Keldysh calculation of the tunneling conductance using the boundary model defined above and introducing Keldysh contour, see Appendix~\ref{App:A}. After Keldysh rotation of the fields $\Theta_{cl/q}=\Theta_+\pm\Theta_-$~\cite{Kamenev}, the corresponding real-time action is given by
\begin{align}
&Z_C=\int D \Theta_{cl} D \Theta_{q}\exp\left[\frac i 4 S_0[\Theta_{cl},\Theta_{q}]\right.\\
&\left.\!-\!2i\lambda_M\!\int\! dt \!\sin\!\left(\frac{2\Theta_{cl}(t)\!+\!a_{cl}(t)}{4}\!\right)\!\sin\!\left(\frac{2\Theta_{q}(t)\!+\!a_{q}(t)}{4}\!\right)\! \right],\nonumber
\end{align}
where $S_0[\Theta_{cl},\Theta_{q}]$ is defined as
\begin{align}
S_0[\Theta_{cl},\Theta_{q}]=\int \frac{d\omega}{2\pi} (\Theta_{\rm cl},\Theta_{\rm q})_{\omega}{\bf G_{\Theta}^{-1}}(\omega)
(\Theta_{\rm cl},\Theta_{\rm q})_{-\omega}
\end{align}
with the Green's function ${\bf G_{\Theta}^{-1}}(\omega)$ being the $2\times 2$ matrix in Keldysh space, see Eq.\eqref{eq:Greens_matrix}.
The current through the junction can be obtained using the following relation
\begin{align}\label{eq:current_def}
\langle I_{cl}(t)\rangle=2i e\frac{\delta Z}{\delta a_q(t)}\left|_{a_q(t)\rightarrow 0}.\right.
\end{align}
By explicitly taking the functional derivative and computing the response function to the lowest non-vanishing order in $\lambda_M$, one finds that current is given by
\begin{align}
\langle I_{cl}(t)\rangle=-2ie\lambda_M^2\int_{-\infty}^{\infty} dt'  C_{\Theta}(t-t')
\end{align}
where the connected correlation function $C_{\Theta}(t-t')$ reads
\begin{widetext}
\begin{align}\label{eq:C_corr_t}
C_{\Theta}(t-t')&=\left \langle \sin\left[\frac{2\Theta_{cl}(t)+a_{cl}(t)}{4}\right]\sin\left[\frac{2\Theta_{cl}(t')+a_{cl}(t')}{4}\right] \cos\left[\frac{\Theta_{q}(t)}{2}\right]\sin\left[\frac{\Theta_{q}(t')}{2}\right] \right \rangle\\
&=\frac{1}{4}\exp\left[-\frac{1}{8}\langle[\delta \Theta^{(-)}_{cl}]^2\rangle\right]\sin\left(\frac{a_{cl}(t)-a_{cl}(t')}{4}\right)\left(\sinh\left[\frac{\langle\delta \Theta^{(-)}_{cl}\delta \Theta^{(-)}_{q}\rangle}{4}\right]-\sinh\left[\frac{\langle\delta \Theta^{(-)}_{cl}\delta \Theta^{(+)}_{q}\rangle}{4}\right]\right)\nonumber
\end{align}
\end{widetext}
with $\delta \Theta^{(\pm)}=\Theta(t)\pm \Theta(t')$. Here the average is taken with respect to $S_0[\Theta_{cl},\Theta_{q}]$.

We now focus on transport near {\it normal reflection} fixed point which describes high temperature/voltage regime. We will assume throughout this section that the thermal coherence length $L_T\equiv v/T$ is shorter than the length of the LL ($L_T\ll L$) in which case one can ignore effect of the leads, see Sec.\ref{sec:leads} for details. With these approximations, the correlation functions $\langle\delta \Theta_{cl}^2\rangle$ and $\langle\delta \Theta^{(-)}_{cl}\delta \Theta^{(\pm)}_{q}\rangle$ can be calculated analytically, see Eqs.\eqref{eq:corrtheta}. Next, given that in the dc limit $a_{cl}(t)-a_{cl}(t')$ is antisymmetric in $t-t'$, the expression~\eqref{eq:C_corr_t} can be simplified by dropping the antisymmetric part of the integrand. The corresponding expression for the tunneling current is given by
\begin{align}
\langle I_{cl}(t)\rangle&=\frac{e\lambda_M^2}{2}\int_{-\infty}^{\infty} dt' \exp\left[-\frac{\langle[\delta \Theta^{(-)}_{cl}]^2\rangle}{8}\right]\\
&\times\sin\left(\frac{a_{cl}(t)\!-\!a_{cl}(t')}{4}\right)\sin\left[\frac{\pi}{2K}{\rm sgn}(t\!-\!t')\right]\nonumber\\
&=\frac{e\lambda_M^2}{\pi T}\left(\frac{\pi T}{\Lambda}\right)^{\frac{1}{K}} \!\!\int_{0}^{\infty} dx \frac{\sin\left(\frac{eV x}{4\pi T}\right)}{\sinh^{\frac{1}{K}}(x)}\sin\left(\frac{\pi}{2K}\right)\label{eq:integral_theta}.
\end{align}
Here $\Lambda$ is an ultraviolet cutoff. As expected, the expression for current in the dc limit is independent of time. After evaluating the integral~\eqref{eq:integral_theta}, the dc current through the junction is given by
\begin{align}\label{eq:current_high}
\!&I_{\rm dc}(V, T)\!=i\frac{e \lambda_M^2}{4\pi T}\left(\frac{\pi T}{\Lambda}\right)^{\frac{1}{K}} 2^{\frac1 K}\Gamma\left(1 \!-\! \frac{1}{K}\right) \sin\left(\frac{\pi}{2K}\right)\nonumber\\
&\times\!\left( \frac{\Gamma\left(\frac{1}{2K}\! +\! i \frac{\tilde V}{2}\right)}{\Gamma\left(1 \!-\! \frac{1}{2K} \!+\! i \frac{\tilde V}{2}\right)} \!-\!  \frac{\Gamma\left(\frac{1}{2K} \!-\! i \frac{\tilde V}{2}\right)}{\Gamma\left(1 \!-\! \frac{1}{2K} \!-\! i \frac{\tilde V}{2}\right)}\right)
\end{align}
where $\tilde V=\frac{eV}{4\pi T}$ and $\Gamma(x)$ is the gamma function. Eq.\eqref{eq:current_high} is the main result of this section as it allows one to study current as a function of an arbitrary ratio of two energy scales $eV$ and $T$. The $I-V$ curve non-trivially depends on the interaction parameter $K$ and in general (i.e. for $K\neq 1$) is non-Ohmic.

One can now obtain the tunneling conductance $G=dI/dV$ in different limits. We first discuss the $eV\ll T$ case. After straightforward manipulations, the dc conductance reads
\begin{align}\label{eq:conductance_high1}
G&=\frac{e^2}{h}\frac{\lambda_M^2}{\Lambda ^2}\left(\frac{\pi T}{\Lambda}\right)^{\frac{1}{K}-2}f_1(K),\\
f_1(K)&=\frac{\sqrt{\pi^3}}{8}\cos\left(\frac{\pi}{2K}\right)\Gamma\!\left(\frac{1}{2} \!-\! \frac{1}{2K}\right) \Gamma \!\left(\frac{1}{2K}\right).\nonumber
\end{align}
As one can see, the conductance $G$ is essentially determined by the scaling dimension of $\lambda_M$ which follows from the RG flow. The prefactor $f_1(K)$ is finite in the non-interacting limit $f_1(K=1)\approx\pi^3/8$. Thus, the result~\eqref{eq:conductance_high1} is consistent with the non-interacting limit.

The non-linear tunneling conductance in the limit $eV\gg T$ reads
\begin{align}\label{eq:conductance_high2}
G\!=\frac{e^2}{h}\frac{\lambda_M^2}{\Lambda^2}\left(\frac{eV}{\Lambda}\right)^{\frac{1}{K}\!-\!2}f_2(K)\\
f_2(K)=-\frac{4\pi \sin\left(\frac{\pi}{K}\right)}{2^{\frac 2 K}}\Gamma\!\left(\!2\!-\!\frac{1}{K}\!\right).
\end{align}
Similarly to the previous case, the power-law dependence of $G\propto V^{\frac{1}{K}\!-\!2}$ is consistent with the scaling theory. However, the pre-factor $f_2(K)\approx \pi^2(1-K)$ vanishes in the non-interacting limit $K=1$ and, thus, the conductance is zero at this order of perturbation theory in $\lambda_M$ (up to corrections $O(\exp[-eV/T])$). This is consistent with the non-interacting results discussed in the previous section where we showed that $G\propto V^{-2}$, cf. Eq.\eqref{eq:conductance_noni_V}. Our result, however, indicates that as soon as we add interparticle interactions, the non-linear conductance scales as $G\propto V^{\frac{1}{K}\!-\!2}$ in agreement with the scaling theory.  Moreover, one can notice that the non-linear conductance~\eqref{eq:conductance_high2} changes sign across $K=1$ which is associated with the change of the local density of states dependence $\rho_{\rm LDOS}(x=0,\omega)\propto \omega^{1/K-1}$~\cite{Fidkowski2012}.

The results for the tunneling conductance \eqref{eq:conductance_high1} and \eqref{eq:conductance_high2} are valid as long as $T\gg \Lambda^*$ and $eV\gg \Lambda^*$, respectively. Here the scale beyond which the perturbation theory breaks down is defined as $\Lambda^*\sim \Lambda (\lambda_M/\Lambda)^{2/(2-K^{-1})}$. Thus, in order to access the low temperature/voltage behavior of the conductance, we need to compute transport starting from the low-energy {\it perfect Andreev reflection} fixed point.

\begin{widetext}
\begin{table*}[tbp]
\begin{tabular}{|c|c|}
\hline
$eV\ll T$ & $eV\gg T$ \\
\hline
\phantom{blah} &   \phantom{blah}\\
$G=\frac{e^2}{h}\frac{\lambda_M^2}{\Lambda ^2}\left(\frac{\pi T}{\Lambda}\right)^{\frac{1}{K}-2}f_1(K)$ \,\,\, \,\,\,& $G=\frac{e^2}{h}\frac{\lambda_M^2}{\Lambda^2}\left(\frac{eV}{\Lambda}\right)^{\frac{1}{K}\!-\!2}f_2(K)$ \,\,\,\,\,\,\,\,\,\\
\phantom{blah} &   \phantom{blah}\\
\hline
\end{tabular}
\caption{Tunneling conductance near {\it perfect normal reflection} fixed point, see Eqs.\ref{eq:conductance_high1} and \eqref{eq:conductance_high2} for details.}
\end{table*}
\end{widetext}

\subsection{Conductance near {\it prefect Andreev reflection} fixed point.}\label{sec:andreev}

In this section, we discuss transport properties of the TP SC/LL junction in the low-energy limit, i.e. near the {\it perfect Andreev reflection} fixed point. For pedagogical reasons, we consider here a simplified setup where $L\rightarrow \infty$ and relegate the discussion of the leads to the next section. We assume that the system renormalizes to the {\it perfect Andreev reflection} fixed point (i.e. the field $\Theta$ is pinned), and the effective low-energy theory is determined by the following  imaginary-time boundary action\cite{Fidkowski2012}:
\begin{align}
S=\frac{1}{2\pi K}\int \frac{d\omega}{2\pi} |\omega||\Phi(\omega)|^2-\lambda_N \int d\tau \cos(2\Phi).
\end{align}
The second term here represents the leading irrelevant perturbation around this fixed point which promotes normal reflection at the junction.

The corresponding partition function for the boundary theory on a Keldysh contour reads (see Appendix~\ref{App:A} for details)
\begin{align}\label{eq:partition_phi}
Z_C&=\int D \Phi_{cl} D \Phi_{q}\\
&\times\exp\left[\frac i 4 S_0[\Phi_{cl},\Phi_{q}]+i S_V[\Phi_{cl},\Phi_{q}]+iS_{\lambda_N}[\Phi_{cl},\Phi_{q}]\right]\nonumber
\end{align}
where $S_0[\Phi_{cl},\Phi_{q}]$ is defined as
\begin{align}
S_0[\Phi_{cl},\Phi_{q}]=\int \frac{d\omega}{2\pi} (\Phi_{\rm cl},\Phi_{\rm q})_{\omega}{\bf G_{\Phi}^{-1}}(\omega)
(\Phi_{\rm cl},\Phi_{\rm q})_{-\omega}
\end{align}
with ${\bf G_{\Phi}^{-1}}(\omega)$ being the $2\times 2$ matrix defined in Eq.~\eqref{eq:Greens_matrix}. The action $S_V[\Phi_{cl},\Phi_{q}]$ describes coupling to the electric field
\begin{align}\label{eq:SV}
S_V[\Phi_{cl},\Phi_{q}]=- \frac{e}{2 \pi }\int_{-\infty}^\infty d t \left[\Phi_q(t) V_{cl}(t)+\Phi_{cl}(t) V_{q}(t)\right]
\end{align}
with $V(t)$ being the applied voltage. The last term in the exponent of \eqref{eq:partition_phi} represents normal backscattering at the junction
\begin{align}\label{eq:Slambda}
S_{\lambda_N}[\Phi_{cl},\Phi_{q}]=-2\lambda_N\int dt \sin\Phi_{cl}(t)\sin \Phi_{q}(t).
\end{align}
Given that irrelevant coupling $\lambda_N$ is small, one can compute the response function function perturbatively.

The dc tunneling conductance $G$ can be obtained by first computing the current through the junction using Eq.~\eqref{eq:current_def} and then differentiating it with respect to the dc voltage $V$
\begin{align}\label{eq:current_low}
\frac{d\langle I_{cl}(t)\rangle}{dV}=\frac{i e^2}{2\pi^2}\int dt' \langle \partial_t\Phi_{cl}(t)\Phi_{q}(t') \rangle_{V_q\rightarrow 0}
\end{align}
where the average is defined with respect to $S_0+S_V+S_{\lambda_N}$. We now perform a transformation of variables to eliminate the term linear in $V$ in the action by introducing a new variable $\tilde\Phi_{\rm cl}(t)=\Phi_{\rm cl}(t)+\alpha(t)$ with
\begin{align}\label{alpha}
\partial_t \alpha(t)=- e K V.
\end{align}
The correlation function in Eq.\eqref{eq:current_low} can be computed by doing perturbation theory in $\lambda_N$:
\begin{align}
\!\!\partial_t\langle \Phi_{cl}(t)\Phi_{q}(t')\rangle_{\!S_0\!+\!S_{\lambda_N}\!+\!S_V}
\!&\!\approx\!\partial_t\langle \tilde{\Phi}_{cl}(t)\Phi_{q}(t')\rangle_{S_0}\\
\!&\!-\frac 1 2 \partial_t\langle \langle \tilde{\Phi}_{cl}(t)\Phi_{q}(t'){\tilde S}_{\lambda}^2\rangle_{S_0}.\nonumber
\end{align}
Here we used the fact that $\langle \alpha(t)\Phi_{q}(t')\rangle_{S_0}=0$. The zeroth order in the $\lambda_N$ correlation function can be easily computed using Eq.\eqref{eq:retard_greens} yielding the following contribution to the dc conductance
\begin{align}\label{eq:conductance_zero}
G^{(0)}=\frac{2e^2}{h}K.
\end{align}
In the non-interacting limit $K=1$, we recover the previous zero-temperature results~\eqref{eq:conductance_noni_T}. However, in general Eq.~\eqref{eq:conductance_zero} suggests that conductance depends on the Luttinger parameter $K$. This is a rather subtle issue~\cite{Maslov1995, Safi_Schulz, Ponomarenko1995, Oreg1995, Alekseev1996, Chamon1996, DolciniPRL'03, DolciniPRB'05, Thomale} since the result for $G$ depends on the experimental setup. As we show below, the presence of the leads is important for the transport measurements [~\onlinecite{Mourik2012, Das2012, Deng2012, Fink2012}] and the conductance $G=2e^2/h$ regardless of the interactions in the LL.

We now compute the second-order in $\lambda_N$ correction to the conductance and define the following correlation function $C_{\Phi}(t-t')$
\begin{align}\label{eq:C_corr}
\!C_{\Phi}(t\!-\!t')&\!\equiv\!\frac 1 2 \partial_t\langle \tilde{\Phi}_{cl}(t)\Phi_{q}(t'){\tilde S}_{\lambda}^2\rangle_{S_0}.
\end{align}
After some manipulations, one finds that the contribution of the connected diagrams to the correlation function $C_{\Phi}(t-t')$ is given by
\begin{align}
\!&\!C_{\Phi}(t\!-\!t')\!=\!-\!i\frac{\lambda_N^2}{2} \!\int\! dt_1\!d t_2\   e^{-W(t_1\!-\!t_2)}\langle  \partial_t\tilde{\Phi}_{cl}(t)\delta\Phi_{q}^{(+)}\rangle\nonumber\\
&\times \langle \delta{\tilde\Phi}_{cl}^{(-)}\Phi_{q}(t')\rangle\cos[\alpha(t_1)\!-\!\alpha(t_2)]\sin(\kappa_2[t_1\!-\!t_2]).
\end{align}
Here $\delta\Phi^{(\pm)}\equiv \Phi(t_1)\pm\Phi(t_2)$; the functions $W(t)$ and $\kappa_2(t)$ are defined as
\begin{align}
W(t_1-t_2)&=\frac{1}{2}\langle [\delta\tilde\Phi^{(-)}_{cl}]^2\rangle \\
i\kappa_2(t_1-t_2)&=\langle\delta\tilde\Phi^{(-)}_{cl}\delta\Phi^{(+)}_{q}\rangle.
\end{align}
The analytical expressions for these functions are given in Appendix~\ref{App:B}.
After straightforward manipulations, we arrive at
\begin{align}\label{eq:C_corr2}
C_{\Phi}(t-t')=-i\lambda_N^2 \int \frac{d\omega}{2\pi}e^{-i\omega(t-t')}\omega [\langle \tilde{\Phi}_{cl}\Phi_{q}\rangle_{\omega}]^2 F(\omega)
\end{align}
where the function $F(\omega)$ reads
\begin{align}\label{eq:functionF0}
F(\omega)&=-\int_{-\infty}^{\infty}d(t_1-t_2)\sin[\omega (t_1-t_2)]e^{-W(t_1-t_2)}\nonumber\\
&\times\cos[\alpha(t_1)-\alpha(t_2)]\sin [\kappa_2(t_1-t_2)].
\end{align}
According to Eq.\eqref{alpha}, the function $\alpha(t)$ depends linearly on $t$ in the dc limit. Using Eqs.\eqref{eq:W_corr} and \eqref{eq:kappa2}, the function $F(\omega)$ is simplified to
\begin{align}\label{eq:functionF}
F(\omega)&=\frac{2}{\pi T}\left(\frac{\pi T}{\Lambda}\right)^{4K} \int_{0}^{\infty}dx \frac{\sin\left(\frac{\omega}{\pi T}x\right)}{\sinh^{4K}(x)}\nonumber\\
&\times\sin(2\pi K)\cos\left[\frac{eVK}{\pi T} x\right].
\end{align}
 One can notice that the integral~\eqref{eq:functionF} is UV-divergent in the regime of interest, i.e. at $K>1/2$. This is a common issue, often appearing in boundary sine-Gordon quantum impurity problems~\cite{Affleck1993,Saleur1999} when one considers flowing “back” from the IR fixed point.  One way to proceed here is to employ dimensional regularization~\cite{Saleur1999}, i.e. we compute the integral in the domain of $K$, where it is well-defined, and then analytically continue to $K>1/2$. 

At small frequencies $\omega \ll T$, one can simplify the expression for $F(\omega)$
\begin{align}
F(\omega)\approx \left\{\begin{array}{cc}
\frac{2\omega}{\Lambda^2}\left(\frac{\pi T}{\Lambda}\right)^{4K-2} g_1(K), & eV/T \ll 1\\& \, \\
\frac{2 \omega}{\Lambda^2}\left(\frac{eVK}{\Lambda}\right)^{4K-2}g_2(K), &  eV/T \gg 1.  \\
                                                                                  \end{array}\right.
\end{align}
Here the dimensionless functions $g_1(K)$ and $g_2(K)$ are given by
\begin{align}\label{eq:g12}
g_1(K)&= -2^{4K-1} \pi \cos(2\pi K)\frac{\Gamma(-4K) \Gamma(1 + 2K)}{
 \Gamma(1 - 2K)}\\
g_{2}(K)&=-2K(4K-1)\Gamma(-4K)\sin(4\pi K).
\end{align}
In the non-interacting limit $K=1$, these functions are finite:
 $g_1(K=1)\approx \pi/3$ and $g_2(K=1)\approx \pi/4$.

Taking into account Eqs.\eqref{eq:current_low}, \eqref{eq:C_corr2} and \eqref{eq:functionF}, the corrections to the dc conductance can be now calculated in both limits:
\begin{align}\label{eq:conductance_final_infinite}
\!\delta G^{(2)}\!=\!\left \{
\begin{array}{cc}
  -\frac{e^2}{h}\frac{8\lambda_N^2 K^2}{\pi\Lambda^2}\left(\frac{\pi T}{\Lambda}\right)^{(4K-2)}g_1(K)  & eV\ll T \\&\\
  -\frac{e^2}{h}\frac{8\lambda_N^2 K^2}{\pi\Lambda^2}\left(\frac{eV K}{\Lambda}\right)^{(4K-2)}g_{2}(K) & eV\gg T.
\end{array}
\right.
\end{align}
As one can see, the scaling of the dc conductance is consistent with the non-interacting results discussed in Sec.\ref{sec:qual}.

\subsection{Conductance near {\it prefect Andreev reflection} fixed point: effect of the leads}\label{sec:leads}

In this section, we consider a finite-length LL and discuss the effect of the leads on the conductance near the low-energy {\it perfect Andreev reflection} fixed point. The experimental setup we consider here is shown in Fig.\ref{Fig:setup}. We will use here an inhomogeneous Luttinger-liquid model and introduce an inhomogeneous interaction parameter $K(x)$ and velocity $v(x)$, which adiabatically interpolate between the nanowire and lead, see Ref.~\cite{Maslov1995} for more details. Since electrons in the lead form a Fermi liquid, which for the present purposes is treated as a non-interacting Fermi gas, we assume here that $K_L=1$. In this case, the naive RG analysis does not work due to the spatial dependence of $K(x)$ and $v(x)$.

We, therefore, pursue a different route here and first compute the Green's function for the LL liquid coupled to the leads at $x=L/2$ and the TP SC at $x=0$.
At the {\it perfect Andreev reflection} fixed point, $\theta(x=0,t)$ is pinned, and, thus, an appropriate boundary condition is $\partial_t \theta(x=0,t)=0$. In terms of the fluctuating variable $\phi$, this is equivalent to $\partial_x \phi(x=0,t)=0$. Similarly, one can define boundary conditions at $x=L/2$, see Appendix~\ref{App:B3}. The real-space Green's function for the Luttinger liquid coupled to the leads can be found by solving
\begin{align}\label{eq:wave}
\left\{\frac{\omega^2}{K(x)v(x)}+\partial_x\left(\frac{v(x)}{K(x)}\partial_x\right)\right\}{\cal G}_\omega(x,x')=\delta(x-x')
\end{align}
where we take the simplest ansatz for the spatially-inhomogeneous velocity $v(x)$ and Luttinger parameter $K(x)$~\cite{Maslov1995}
\begin{align}
v(x)&=\left\{\begin{array}{cc}
                                                                                   v &  x\leq L/2  \\
                                                                                    \upsilon' &  x>L/2
                                                                                  \end{array}
\right.\\
K(x)&=\left\{\begin{array}{cc}
                                                                                   K_W &  x\leq L/2  \\
                                                                                    K_L &  x>L/2.
                                                                                  \end{array}
\right.
\end{align}
The straightforward solution of the above wave equation $G_{\omega}(x,x')$ at the origin $x=x'=0$ (see Appendix~\ref{App:B3} for details) reads
\begin{align}\label{eq:GreensL}
\!{\cal G}_{\omega}(0,0)\!=\!\frac{K_w}{i\omega}\cdot\frac{K_w\!+\!K_L\!+\!(K_L\!-\!K_w)e^{i\omega L/v}}{K_w\!+\!K_L\!-\!(K_L\!-\!K_w)e^{i\omega L/v}}.
\end{align}
One can notice that there is an additional energy scale in the problem, $E_L=v/L$, which is associated with time of propagation through the nanowire to the leads.

Using Eq.\eqref{eq:GreensL}, it is straightforward to obtain the modified Keldysh action in the presence of leads at the Andreev fixed point.
\begin{align}\label{eq:S0L}
{\tilde S}_0[\Phi_{cl},\Phi_{q}]&=\int \frac{d\omega}{2\pi} (\Phi_{\rm cl},\Phi_{\rm q})_{\omega}{\bf \tilde{G}_{\Phi}^{-1}}(\omega)
(\Phi_{\rm cl},\Phi_{\rm q})_{-\omega}
\end{align}
where the Green's functions are given by
\begin{align}\label{eq:GReensL0}
{\cal \tilde G}_{\Phi}^{R/A}(\omega)&=\mp\frac{\pi i K_w}{\omega\pm i\delta}\cdot\frac{K_w\!+\!K_L\!+\!(K_L\!-\!K_w)e^{\pm i\omega L/v}}{K_w\!+\!K_L\!-\!(K_L\!-\!K_w)e^{\pm i\omega L/v}}\\
{\cal \tilde G}_{\Phi}^{K}(\omega)&=- \frac{2\pi i K_w}{\omega}\coth\left(\frac{\omega}{2T}\right)\\
&\times\frac{2K_L K_w}{K_L^2+K_w^2-(K_L^2-K_w^2)\cos\left(\frac{\omega L}{v}\right)}.\nonumber
\end{align}

We now compute tunneling conductance in the low-energy limit. The presence of a new scale in the problem $E_L$ modifies the flow to the IR fixed point. However, Majorana coupling is a relevant perturbation for a wide range of the Luttinger parameter $K$ which includes the non-interacting value $K=1$ in the semi-infinite lead. Therefore, the system eventually flows to the Andreev fixed point as in the homogeneous case (at least for small mismatch of the Luttinger parameters $K_w$ and $K_L$).  As in Sec.~\ref{sec:andreev}, we calculate tunneling conductance by including the leading irrelevant perturbation near Andreev fixed point corresponding to fermion backscattering processes.

The full Keldysh action is $S_C=\tilde S_0+S_V+S_{\lambda_N}$ where $\tilde S_0$,$S_V$ and $S_{\lambda_N}$ are defined in Eqs.\eqref{eq:S0L}, \eqref{eq:SV} and \eqref{eq:Slambda}, respectively. Using Eq.\eqref{eq:current_low}, we first compute the conductance to zeroth order in $\lambda_N$. It is straightforward to show that the dc conductance $G^{(0)}$ (cf. \ref{eq:conductance_zero}) is given by
\begin{align}
G^{(0)}=i\frac{2e^2}{\pi h}\omega {\cal \tilde G}^R_{\Phi}(\omega)|_{\omega \rightarrow 0}=\frac{2e^2}{h}K_L.
\end{align}
Note that the conductance depends on $K_L=1$ rather than $K_w$, which is consistent with the previous works on a similar problem~\cite{Maslov1995, Safi_Schulz, Ponomarenko1995, Alekseev1996, Chamon1996}. Thus, in a typical setup for measuring dc transport, see Fig.\ref{Fig:setup}, the {\it perfect Andreev reflection} fixed point corresponds to a quantized conductance in units of $2e^2/h$ even in the presence of the interactions in the nanowire.

We discuss next the corrections to conductance at finite temperature and voltage. In order to compute the corrections to the lowest non-vanishing order in $\lambda_N$, we first shift the field $\Phi_{cl}(t)\rightarrow \Phi_{cl}(t)+\alpha(t)$  and eliminate the $S_V$ term in the Keldysh action. In the case of an applied dc voltage $V$, the function $\alpha(t)$ is given by
$$\alpha(t)=2eV\int \frac{d\omega}{2\pi}\delta(\omega){\cal G}_{\Phi}^{R}(\omega)e^{- i \omega t}.$$ Using Eq.\eqref{eq:GReensL0}, one finds that $\partial_t \alpha(t)=-eVK_L$. The rest of the calculation proceeds exactly as in the homogeneous LL case discussed in Sec.\ref{sec:andreev}. The only difference here is that the correlation function \eqref{eq:C_corr} is now averaged with respect to the action $\tilde{S}_0$ \eqref{eq:S0L}:
\begin{align}\label{eq:C_corr2L}
\tilde{C}_{\Phi}(t-t')=4i\lambda_N^2 \int \frac{d\omega}{2\pi}e^{-i\omega(t-t')}\omega [{\cal \tilde G}^R_{\Phi}(\omega)]^2 {\tilde F}(\omega).
\end{align}
The function $\tilde{F}(\omega)$ is given by
\begin{align}\label{eq:functionF0L}
\tilde{F}(\omega)&=-\int_{-\infty}^{\infty}d\tau\sin(\omega \tau)e^{-{\tilde W}(\tau)}\cos[eV_L \tau]\sin[\tilde{\kappa}_2(\tau)]
\end{align}
where $\tilde W(\tau)$ and $\tilde \kappa_2(\tau)$ are defined in Eqs.\eqref{eq:W_corr_leads2} and \eqref{eq:kappa2leads}, respectively. In general, $\tilde W(\tau)$ and $\tilde \kappa_2(\tau)$ are complicated functions of $K_w$ and $K_L$, and the integral~\eqref{eq:functionF0L} has to be evaluated numerically. One can also evaluate these functions analytically in the limit of small $\delta K \equiv K_w-K_L$, see Eqs.\eqref{eq:W_corr_leads2} and \eqref{eq:kappa2leads}, which allows one to compute the conductance in the limit of weak electron-electron interactions in the nanowire. Henceforth, we assume that $|\delta K|\ll 1$.

The function $\tilde{F}(\omega)$ also depends on the ratio between $T, eV$, and $E_L$, i.e. the main contribution to the integral~\eqref{eq:functionF0L} can come from the interval $\tau < L/v$ or $\tau > L/v$ subject to the infrared (IR) cutoff determined by ${\rm max}(eV,T)$. If the wire length is much larger than the thermal coherence length $L_{T}$, the correlation function $\tilde{F}(\omega)$ is approximately given by
\begin{align}\label{eq:F11}
\tilde{F}(\omega)&\approx \frac{2}{\pi T}\left(\frac{\pi T}{\Lambda}\right)^{4K_w} \int_{0}^{\frac{\pi T L}{v}}dx \frac{\sin\left(\frac{\omega}{\pi T}x\right)}{\sinh^{4K_w}(x)}\nonumber\\
&\times \sin(2\pi K_w)\cos\left[\frac{eVK_L}{\pi T} x\right].
\end{align}
At small frequencies $\omega \ll T$, one finds that
\begin{align}
\tilde{F}(\omega)\approx \left\{\begin{array}{cc}
                                                                                   \frac{2\omega}{\Lambda^2}\left(\frac{\pi T}{\Lambda}\right)^{4K_w-2} g_1(K_w), & eV/T \ll 1\\&\\
                                                                                    \frac{2 \omega}{\Lambda^2}\left(\frac{eVK_L}{\Lambda}\right)^{4K_w-2}g_2(K_w), &  eV/T \gg 1
                                                                                  \end{array}\right.
\end{align}
where the functions $g_1(K)$ and $g_2(K)$ are defined in Eqs.\eqref{eq:g12}.

In the opposite limit $L\ll L_{T}$, the function $\tilde{F}(\omega)$ is approximately given by
\begin{align}
&\tilde{F}(\omega)\approx\frac{2}{\pi T}\left(\frac{\pi T}{\Lambda}\right)^{4K_L} \left(\frac{v}{L \Lambda}\right)^{4\delta K}\\
&\times\int_{\frac{\pi T L}{v}}^{\infty}dx \frac{\sin\left(\frac{\omega}{\pi T}x\right)}{\sinh^{4K_L}(x)}\sin(2\pi K_L)\cos\left[\frac{eVK_L}{\pi T} x\right].\nonumber
\end{align}
In the low-frequency limit $\omega \ll T$, the expression for $\tilde{F}(\omega)$ simplifies to
\begin{align}
\!\!\tilde{F}(\omega)\!\approx\! \left\{\begin{array}{cc}
                                                                                  \frac{2\omega}{\Lambda^2}\left(\frac{v}{L \Lambda}\right)^{4\delta K}\left(\frac{\pi T}{\Lambda}\right)^{4K_L-2}g_1(K_L),  & \frac{eV}{T} \ll 1\\&\\
                                                                                   \frac{2 \omega}{\Lambda^2}\left(\frac{v}{L \Lambda}\right)^{4\delta K}\left(\!\frac{eVK_L}{\Lambda}\!\right)^{4K_L-2}g_{2}(K_L), & \frac{eV}{T} \gg 1.
                                                                                  \end{array}\right.
\end{align}
Here we included only leading terms in the expansion in $|\delta K|\ll 1$. In the case of $eV \gg v/L \gg T$, we also find the subleading oscillatory corrections proportional to $\cos(eV L/v)$. These oscillations originates from the interference effects of plasmon modes in the nanowire due to the backscattering from the TP-SC/LL and lead/LL boundaries. The reflection coefficient at the lead/LL boundary depends on $\delta K$~\cite{Safi'97}. We note that the RG analysis captures only leading order corrections to the conductance. The increase of the mismatch parameter $|\delta K|$ leads to the enhancement of the backscattering at the lead/LL boundary, and the RG analysis, which does not capture the aforementioned interference effects, will eventually break down. We also note here that there are no such oscillatory corrections in the high-temperature limit $T \gg v/L \gg eV$. This is not surprising since temperature suppresses the interference effects. Indeed, although voltage and temperature provide high energy cutoffs in the RG analysis, they regularize corresponding integrals in a very different way, see Eq.\eqref{eq:F11}. We refer a reader to Refs.~\cite{DolciniPRL'03, DolciniPRB'05} for more details on this point.

Having these caveats in mind, we can now discuss the dc conductance in the presence of leads. We proceed as before and use Eqs.\eqref{eq:current_low},\eqref{eq:C_corr2L} and \eqref{eq:functionF0L} to calculate the corrections to the conductance. As expected, in case of $L \gg L_T$ the presence of the leads is irrelevant, and we obtain the same answer as in the $L\rightarrow \infty$ limit, cf. Eq.~\eqref{eq:conductance_final_infinite} where $K$ is the Luttinger parameter in the wire. On the other hand, if $L\ll L_T$ one finds that the temperature dependence of the corrections to $G$ is governed by the Luttinger parameter of the leads $K_L$. The conductance $G$ in this regime is given by
\begin{widetext}
\begin{align}\label{eq:conductance_final_finite}
G&\approx\frac{2e^2}{h}K_L-\frac{e^2}{h}\frac{8\lambda_N^2 K_L^2}{\Lambda^2}\left(\frac{\pi T}{\Lambda}\right)^{(4K_L-2)}\left(\frac{v}{L \Lambda}\right)^{4\delta K}g_1(K_L)\mbox{ for } eV\ll T\nonumber\\&\\
G\!&\!\approx\frac{2e^2}{h}K_L\!-\!\frac{e^2}{h}\frac{8\lambda_N^2 K_L^2}{\Lambda^2}\left(\frac{eV K_L}{\Lambda}\right)^{(4K_L-2)} \left(\frac{v}{L \Lambda}\right)^{4\delta K} g_{2}(K_L)\mbox{ for } eV\gg T\nonumber
\end{align}
\end{widetext}
where we eventually have to set $K_L=1$. Thus, in this limit temperature/voltage corrections to the tunneling conductance scale similarly as in the non-interacting case. One can notice, however, that $\delta G$ has a non-trivial dependence on the nanowire length $L$ and the Luttinger parameter $K_w$. In the case of repulsive interactions in the LL ($K_w<1$), the correction to the conductance is enhanced by the factor $\left(\frac{v}{L \Lambda}\right)^{4\delta K}$.

\begin{widetext}
\begin{table*}[tbp]
\begin{tabular}{|c|c|c|}
\hline
\phantom{blah} & $L \gg L_T$ & $L \ll L_T$ \\
\hline
\phantom{blah} & \phantom{blah} &   \phantom{blah}\\
$eV\ll T$ & $\delta G^{(2)}\approx -\frac{e^2}{h}\frac{8\lambda_N^2 K_L^2}{\Lambda^2}\left(\frac{\pi T}{\Lambda}\right)^{(4K_w-2)}g_1(K_w)$ \,\,\, \,\,\,& $\delta G^{(2)}\approx-\frac{e^2}{h}\frac{8\lambda_N^2 K_L^2}{\Lambda^2}\left(\frac{\pi T}{\Lambda}\right)^{(4K_L-2)}\left(\frac{v}{L \Lambda}\right)^{4\delta K}g_1(K_L)$ \,\,\,\,\,\,\,\,\,\\
\phantom{blah} & \phantom{blah} &   \phantom{blah}\\
\hline
\phantom{blah} & \phantom{blah} &   \phantom{blah}\\
$eV\gg T$ & $\delta G^{(2)}\approx-\frac{e^2}{h}\frac{8\lambda_N^2 K_L^2}{\Lambda^2}\left(\frac{eV K_L}{\Lambda}\right)^{(4K_w-2)}g_{2}(K_w)$ & $\delta G^{(2)}\approx-\frac{e^2}{h}\frac{8\lambda_N^2 K_L^2}{\Lambda^2}  \left(\frac{eV K_L}{\Lambda}\right)^{(4K_L-2)} \left(\frac{v}{L \Lambda}\right)^{4\delta K} g_{2}(K_L)$ \\
\phantom{blah} & \phantom{blah} &   \phantom{blah}\\
\hline
\end{tabular}
\caption{Corrections to the tunneling conductance near {\it perfect Andreev reflection} fixed point, see Eq.\eqref{eq:conductance_final_infinite} and
\eqref{eq:conductance_final_finite} for more details. Here $L_T\equiv v/T$ is the thermal coherence length.}
\end{table*}
\end{widetext}

\section{Qualitative discussions of main results}\label{sec:mainresults}

We now review our main results for the tunneling conductance in the TP SC/LL junction. At high energies (i.e. near the {\it perfect normal reflection} fixed point), the results for the tunneling conductance $G$ are given in Table I. We find that the temperature and voltage dependence of the conductance are consistent with the scaling theory. However, the pre-factor of the non-linear conductance~\eqref{eq:conductance_high2} vanishes in the non-interacting limit $K_w=1$. This is consistent with the previous results based on the non-interacting calculation, see Sec.~\ref{sec:qual}. Thus, the non-interacting results for the conductance are quite fine-tuned and do not represent the generic situation in the nanowires where electron-electron interactions are inevitably present. We show that the non-linear conductance changes sign across the $K_w=1$ point which is associated with the interaction-induced enhancement/suppression of the local density of states at the junction.

The low-energy results for the tunneling conductance $G$ obtained near the {\it perfect Andreev reflection} fixed point are summarized in Table II. First of all, we showed that in the presence of the leads, the dc tunneling conductance is equal to $2e^2/h$ at zero temperature. This is consistent with the previous studies~\cite{Maslov1995, Safi_Schulz, Ponomarenko1995, Alekseev1996}. In Sec.~\ref{sec:andreev} we studied finite-temperature and finite-voltage corrections to the tunneling conductance and showed that the power-law dependence on $T$ and $V$ is not universal and non-trivially depends on the length of the LL region $L$, see Table II. In the limit of small Luttinger parameter mismatch ($|\delta K| \ll 1$), these results can be understood using a two-step RG analysis discussed below. When the mismatch parameter increases, one should include plasmon interference effects due to the backscattering from lead/LL boundary.

We now focus on small mismatch limit $\delta K \ll 1$ and provide a qualitative argument how to understand leading order corrections to conductance in the presence of the lead. Let us consider the scattering from the boundary at $x=0$ as shown in Fig.\ref{Fig:setup}. In the unfolded picture discussed in Sec.~\ref{sec:qual}, the boundary can be viewed as an effective ``impurity". It is well known that charge density oscillations (i.e. Friedel oscillations) around such an ``impurity" lead to the dynamically-generated effective potential which determines the transmission/reflection probability. The spatial extent of the Friedel oscillations is determined by the thermal coherence length $L_T=v/T$. Thus, at high temperature $L_T \ll L$, the reflection probability from the impurity depends on $v$ and $K_w$, whereas in the opposite limit the effective size of impurity potential becomes larger than the length of the wire, and scattering is governed by the dynamics in the leads characterized by $v'$ and $K_L$. These arguments suggest that one can understand the results~\eqref{eq:conductance_final_finite} using a two-step RG flow of the coupling constant $\lambda_N$. In the homogeneous case (i.e. $L\rightarrow \infty$), the RG equation for $\lambda_N$ is given in Eq.~\eqref{eq:lambda_N} where $K$ is the LL parameter in the nanowire. If the LL length $L$ is finite, there is an additional energy scale in the problem $E_L \equiv v/L$ which separates the two aforementioned regimes. Specifically, in the high-energy (short-length scale) regime $\Lambda > E > E_L$, the flow of the coupling constant $\lambda_N$ is governed by $K_w$. If ${\rm max}\{T,eV\} \gg E_L$, we have to stop the RG flow before it reaches $E_L$. Therefore, the expression for the effective coupling constant at the IR cutoff scale is independent of the nanowire length $L$:
\begin{align*}
\lambda^*_N=\lambda_N(0)\left(\frac{{\rm max}\{T,eV\}}{\Lambda}\right)^{2K_w-1}.
\end{align*}
The correction to the conductance is proportional to $(\lambda^*_N)^2$ and is given by
\begin{align}\label{eq:scaling1}
\delta G \propto -\lambda^2_N(0) \left(\frac{{\rm max}\{T,eV\}}{\Lambda}\right)^{4K_w-2}.
\end{align}
In the case ${\rm max} \{T,eV\}\ll E_L$, the renormalization of $\lambda_N$ is determined by $K_w$ and $K_L$ in the energy intervals $\Lambda > E > E_L$ and $E_L > E > {\rm max} \{eV, T\}$, respectively. A straightforward integration of the RG flow for $\lambda_N$ yields
\begin{align*}
\lambda^*_N=\lambda_N(0)\left(\frac{E_L}{\Lambda}\right)^{2K_w-1}\cdot\left(\frac{{\rm max}\{eV, T\}}{E_L}\right)^{2K_L-1},
\end{align*}
and one finds that the correction to the conductance is given by
\begin{align}\label{eq:scaling2}
\delta G \propto -\lambda^2_N(0) \left(\frac{{\rm max}\{T,eV\}}{\Lambda}\right)^{4K_L-2}\left(\frac{E_L}{\Lambda}\right)^{4(K_w-K_L)}.
\end{align}
The scaling of conductance~\eqref{eq:scaling1} and ~\eqref{eq:scaling2} agrees perfectly with the results obtained using Keldysh formalism in $|\delta K |\ll 1$ limit, see Table II.

Thus, we have shown that the two-step RG approach significantly simplifies the calculation of temperature/voltage corrections to the tunneling conductance and allows one to take into account the effect of the leads in the small mismatch limit $|\delta K| \ll 1$. This approach is quite general and can be easily applied to other quantum impurity problems where the presence of leads is important.

\section{Conclusions}

In this paper, we considered transport properties of a topological superconductor-Luttinger liquid junction and computed the tunneling transport conductance using a real-time Keldysh technique. We have considered a realistic experimental setup~\cite{Mourik2012} and taken into account the effect of the leads within an inhomogeneous Luttinger liquid model. We find that interactions in the Luttinger liquid do not modify the quantization of the zero-bias tunneling conductance at zero temperature but rather lead to a change of the temperature/voltage corrections to the conductance. We show that in some instances the results based on non-interacting models are significantly modified due to the presence of the interactions. Thus, our results have important implications for the tunneling experiments aimed at detecting Majorana zero-energy modes in semiconductor/superconductor heterostructures.


\acknowledgments{It is a pleasure to thank Jason Alicea, Lukasz Fidkowski, Paul Fendley, Leonid Glazman, Nate Lindner, Max Metlitski, Chetan Nayak, Yuval Oreg and especially Matthew Fisher for stimulating discussions. RL wishes to acknowledge the hospitality of the Aspen Center for Physics and support under NSF Grant \#1066293.}


\appendix

\section{Conventions and Notations}\label{App:A}

In this Appendix, we specify our conventions and review basic properties of the Keldysh technique used throughout the main text. We follow closely the Keldysh path integral approach reviewed in Ref.\cite{Kamenev}. We start with the definition of the partition function on a Keldysh contour which reads
\begin{align}
Z=\int D \phi \exp\left( i \int_C dt \phi(t)G^{-1}\phi(t)\right).
\end{align}
The fields on the upper and lower branches of the Keldysh contour are $\phi_+$ and $\phi_-$, respectively. In general, the action $S_C$ can be written as
\begin{align}
S_C=\int_C dt \phi(t)G^{-1}\phi(t)&=\int_{-\infty}^{\infty} dt \phi_+(t)G^{-1}\phi_+(t)\nonumber\\
&-\int_{-\infty}^{\infty} dt \phi_-(t)G^{-1}\phi_-(t).\nonumber
\end{align}
Next, we perform a Keldysh rotation by introducing new fields $\phi_{\rm cl}=\phi_++\phi_-$ and $\phi_{\rm q}=\phi_+-\phi_-$. Without loss of generality, we now consider the action for the finite-length Luttinger liquid, which, after Keldysh rotation, reads
\begin{widetext}
\begin{align}\label{eq:S0}
\!S_C\!=\! \int_{-\infty}^{\infty} \!dt \int_{0}^{L/2} \!dx \!\left[\frac 1 2 (\Pi_{\rm cl},\Pi_{\rm q})\left(\begin{array}{cc}
                                                                                     0 & \partial_t \\
                                                                                     \partial_t & 0
                                                                                   \end{array}
\right)(\phi_{\rm cl},\phi_{\rm q})\!-\!\frac{1}{2\pi}(\Pi_{\rm cl},\Pi_{\rm q})\left(\begin{array}{cc}
                                                                                     0 & \frac{v K \pi^2}{2} \\
                                                                                     \frac{v K \pi^2}{2} & 0
                                                                                   \end{array}
\right)(\Pi_{\rm cl},\Pi_{\rm q})- \frac{v}{2\pi K} \partial_x \phi_{\rm cl}\partial_x \phi_{\rm q} \right].
\end{align}
\end{widetext}
By integrating out $\Pi_{cl}$- and $\Pi_{cl}$-fields in Eq.\eqref{eq:S0}, one finds
\begin{align}\label{eq:partition}
&Z_C=\int D\phi_{\rm cl}D\phi_{\rm q}\exp\left[\frac i 4 S_0[\phi_{cl},\phi_{q}] \right]\\
&S_0[\phi_{cl},\phi_{q}]=\int_{x,t}\int_{x',t'} (\phi_{\rm cl},\phi_{\rm q})_{x,t} {\bf G}^{-1}_{\phi}
(\phi_{\rm cl},\phi_{\rm q})_{x',t'}
\end{align}
where the matrix Green's function ${\bf G}^{-1}_{\phi}$ is defined as
\begin{align}\label{eq:Greens_matrix}
{\bf G}^{-1}_{\phi}=\left(\begin{array}{cc}
                                                                                     0 & G_A^{-1} \\
                                                                                     G_R^{-1} & (G^{-1})_K
                                                                                   \end{array}\right).
\end{align}
The Green's functions are given by
\begin{align}\label{eq:GreensRA_phi}
{\cal G}^{-1}_{A/R}=\frac{-\partial_t^2+v^2 \partial_x^2}{2 \pi K v}\delta(x-x')\delta(t-t')
\end{align}
and $(G^{-1})_{K}$ is defined as~\cite{Kamenev}
\begin{align}\label{eq:GreensK_phi}
({\cal G}^{-1})_{K}={\cal G}_R^{-1}\circ F- F \circ {\cal G}_A^{-1}.
\end{align}
In the translationally-invariant system we find
\begin{align}\label{eq:Greens1}
&{\cal G}^{R/A}(\omega, q)=\frac{2\pi v K}{(\omega\pm i \delta)^2-v^2 q^2}\\
&{\cal G}^{K}(\omega, q)=\coth\left(\frac{\omega}{2T}\right)\left[{\cal G}^R(\omega, q)-{\cal G}^A(\omega, q)\right]\nonumber\\
\!&\!=\!\coth\!\left(\!\frac{\omega}{2T}\!\right)\!\frac{\pi K}{q}\! 2\pi i \left[\!\delta(\omega\!+\!v q)\!-\!\delta(\omega\!-\!v q)\!\right].
\end{align}

The following correlation functions are defined in terms of the Green's functions as
\begin{align}
\langle \phi_{cl}(x,t) \phi_{q}(x',t') \rangle = 2 i {\cal G}^R(xt;x't')\\
\langle \phi_{q}(x,t) \phi_{cl}(x',t') \rangle = 2 i {\cal G}^A(xt;x't')\\
\langle \phi_{cl}(x,t) \phi_{cl}(x',t') \rangle = 2 i {\cal G}^K(xt;x't').
\end{align}

We can now derive the boundary theory by integrating out the bulk modes in Eq.\eqref{eq:partition}. The Keldysh action with the constraint $\Phi(t)=\phi(x=0,t)$ is given by
\begin{align}
\!Z_C[\Phi_{cl},\Phi_q]\!&\!=\!\int D\phi_{cl}D\phi_{q}D\chi_{cl}D\chi_q\exp\!\left[\!\frac i 4 \!S_0[\phi_{cl},\phi_{q}] \right.\nonumber\\
&\!+\!\frac i 2\int dt dx \left(\chi_q(t)\delta(x)[\Phi_{cl}(t)\!-\!\phi_{cl}(x,t)]\right. \nonumber\\
& \left. \left. \!+\chi_{cl}(t)\delta(x)[\Phi_{q}(t)-\phi_{q}(x,t)]\right) \right].
\end{align}
We first integrate out the $\phi(x,t)$ fields and then perform a functional integral over $\chi(t)$. As a result, one arrives at the following boundary action
\begin{align}
&Z_{CB}=\int D \Phi_{cl} D \Phi_{q}\exp\left[\frac i 4 S_0[\Phi_{cl},\Phi_{q}]\right]\\
&S_0[\Phi_{cl},\Phi_{q}]=\\
&\int_{t}\int_{t'} (\Phi_{\rm cl},\Phi_{\rm q})_{t} {\bf G}^{-1}_{\phi}(x=0,t;x'=0,t')
(\Phi_{\rm cl},\Phi_{\rm q})_{t'}.\nonumber
\end{align}
This result is quite general and applies to the inhomogeneous Luttinger liquid model discussed in the main text.

The correlation functions for the $\theta$-field can be straightforwardly obtained by changing $K \rightarrow K^{-1}$ in Eqs.\eqref{eq:GreensRA_phi} and \eqref{eq:GreensK_phi}.

\section{Correlation functions for semi-infinite LL}\label{App:B}

In this section, we derive the expressions for the various correlations functions for the semi-infinite LL case. Due to the translational invariance in the wire, the Green's functions depend on the difference $x-x'$.  Therefore, the retarded and advanced boundary Green's functions become
\begin{align}\label{eq:Green}
{\cal G}_{\Phi}^{R/A}(\Delta t)=- \int \frac{d\omega}{2\pi} \frac{d q}{2\pi}\frac{2\pi v K e^{-\alpha |q|\mp i\omega\Delta t}}{(vq-\omega-i\delta)(vq+\omega+i\delta)}
\end{align}
where $\alpha$ is a cutoff. In the frequency domain, we find
\begin{align}\label{eq:Greens2}
{\cal G}_{\Phi}^{R/A}(\omega)&=\mp \frac{\pi i K}{(\omega \pm i \delta)}\\
{\cal G}_{\Phi}^K(\omega)&=-i\frac{2\pi K}{\omega}\coth\left(\frac{\omega}{2T}\right)
\end{align}
with corresponding real-time Green's functions
\begin{align}\label{eq:retard_greens}
{\cal G}_{\Phi}^{R/A}(\Delta t)&=\mp i \int \frac{d\omega}{2\pi}e^{\mp i\omega\Delta t}\frac{\pi K}{\omega\pm i\delta}\nonumber\\
&=-\pi K \vartheta(\pm \Delta t)
\end{align}
where $\vartheta(t)$ is the Heaviside step function. We note that the equal-time Green's function is equal to zero $G_{\Phi}^{R/A}(0)=0$, as can be seen by performing the frequency integral first in Eq.\eqref{eq:Green}.

Let us now compute several correlation functions that appear in the calculations. The function $W(t)$ appears in the perturbative expansion
\begin{align}\label{eq:W_corr}
&W(t_1-t_2)\equiv\frac{1}{2}\langle(\Phi_{cl}(t_1)-\Phi_{cl}(t_2))^2\rangle\nonumber\\
&=4K\int_{0}^{\infty}dx\frac{1-\cos\left(T(t_1-t_2)x \right)}{x}\coth\left(\frac{x}{2}\right)\nonumber\\
&\!=\!8K\int_{0}^{\infty}dx\left[1\!-\!\cos\left(T(t_1-t_2)x\right)\!\right]\!\sum_{n=-\infty}^{\infty}\frac{\exp[-\frac n \alpha]}{(2\pi n)^2\!+\!x^2}\nonumber\\
&\!=\!2K\log\left(\frac{\Lambda^2}{\pi^2T^2}\!\sinh^2[\pi T(t_1\!-\!t_2)]\right)
\end{align}
where $\alpha=\Lambda/T$ with $\Lambda$ being an UV cutoff. Also, we have assumed here that $|t_1-t_2| \gg \Lambda^{-1}$. As can be easily shown, the correlation function $\langle(\Phi_{cl}(t_1)+\Phi_{cl}(t_2))^2\rangle$ is divergent and, thus, the terms in the perturbative expansion proportional to $\exp[-\frac 1 2 \langle(\Phi_{cl}(t_1)+\Phi_{cl}(t_2))^2\rangle]$ vanish.

The correlation functions $\kappa_1(t_1-t_2)$ and $\kappa_2(t_1-t_2)$ are defined as
\begin{align}\label{eq:kappa2}
i\kappa_1(t_1-t_2)&=\langle\delta \Phi^{(-)}_{cl}\delta \Phi_q^{(-)}\rangle\nonumber\\
&=2\pi i K \left(\vartheta(t_1-t_2)+\vartheta(t_2-t_1)\right)\nonumber\\
&= 2\pi i K\\
i\kappa_2(t_1-t_2)&=\langle\delta \Phi^{(-)}_{cl}\delta \Phi_q^{(+)}\rangle
=-2\pi i K {\rm sgn} (t_1-t_2)
\end{align}
where $\delta \Phi^{(\pm)}=\Phi(t_1) \pm \Phi(t_2)$.

The correlation functions for the $\theta$-field can be similarly obtained. Using the duality between the $\theta$ and $\phi$ variables, one finds
\begin{align}\label{eq:corrtheta}
\frac 1 2 \langle[\delta \Theta^{(-)}_{cl}]^2\rangle&=\frac{2}{K}\log\left(\frac{\Lambda^2}{\pi^2T^2}\!\sinh^2[\pi T(|t_1\!-\!t_2|)]\right)\\
\langle\delta \Theta^{(-)}_{cl}\delta \Theta^{(-)}_{q}\rangle&=\frac{2\pi i}{K}\\
\langle\delta \Theta^{(-)}_{cl}\delta \Theta^{(+)}_{q}\rangle&=-\frac{2\pi i}{K}{\rm sgn}(t_1-t_2).
\end{align}

\section{Green's function for LL coupled to the lead.}\label{App:B3}
We now consider the effect of the leads and derive the real-time Green's function for a LL of length $L/2$ coupled to the metallic lead. We first compute the Green's function for an infinite system and then use mirror image method to impose an appropriate boundary condition at $x=0$. We will use here the inhomogeneous LL model~\cite{Maslov1995} and first solve the following wave equation for the Green's function in the frequency-position domain
\begin{align}
\left\{\frac{\omega^2}{K(x)v(x)}+\partial_x\left(\frac{v(x)}{K(x)}\partial_x\right)\right\}{\cal G}_\omega(x,x')=\delta(x-x').
\end{align}
We take the inhomogeneous velocity $v(x)$ and Luttinger parameter $K(x)$ to be
\begin{align}
v(x)&=\left\{\begin{array}{cc}
                                                                                   v &  |x|\leq L/2  \\
                                                                                    \upsilon' &  |x|>L/2
                                                                                  \end{array}
\right.\\
K(x)&=\left\{\begin{array}{cc}
                                                                                   K_w &  |x|\leq L/2  \\
                                                                                    K_L &  |x|>L/2.
                                                                                  \end{array}
\right.
\end{align}
Since $K(x)$ and $v(x)$ are constants at each point in space, the Green's function takes the general form
\begin{align}
{\cal G}_\omega(x,x')=A(x')e^{i\frac\omega v x}+B(x')e^{-i\frac\omega v x}.
\end{align}
The Green's function should satisfy continuity at $x=\pm L/2$ and $x=x'$ as well as the following boundary conditions at $x=-L/2, x',$ and $L/2$:
\begin{align}
\frac{v(x)}{K(x)}\partial_x {\cal G}_\omega (x,x')\bigg|_{x=\pm L/2}&=0\\
\frac{v(x)}{K(x)}\partial_x {\cal G}_\omega (x,x')\bigg|_{x=x'+0^-}^{x=x'+0^+}&=1.
\end{align}
We are ultimately looking for the solution for the Green's function in the wire and henceforth assume $|x'|\leq L/2$. This divides our space into 4 regions, where we must match the 6 boundary conditions defined above
\begin{align}
{\cal G}^{(\mathrm{I})}_\omega(x,x')&=B_1(x')e^{-i\frac\omega v x}& x < - L/2\nonumber\\
{\cal G}^{(\mathrm{I})}_\omega(x,x')&=A_2(x')e^{i\frac\omega v x}+B_2(x')e^{-i\frac\omega v x}& -L/2\leq x <x'\nonumber\\
{\cal G}^{(\mathrm{II})}_\omega(x,x')&=A_3(x')e^{i\frac\omega v x}+B_3(x')e^{-i\frac\omega v x}& x'\leq x \leq L/2\nonumber\\
{\cal G}^{(\mathrm{III})}_\omega(x,x')&=A_4(x')e^{i\frac\omega v x}& x>L/2.\nonumber
\end{align}
Solving for the coefficients $A_i$ and $B_i$, we find the Green's function at $|x|\leq L/2$.
\begin{widetext}
\begin{align}
{\cal G}_\omega(x,x')=\frac{i K_w}{\omega}\frac{e^{iL/L_\omega}\kappa_-^2e^{-i|x-x'|/L_\omega}+e^{-iL/L_\omega}\kappa_+^2e^{i|x-x'|/L_\omega}+2\kappa_-\kappa_+\cos (|x+x'|/L_\omega)}{e^{iL/L_\omega}\kappa_-^2-e^{-iL/L_\omega}\kappa_+^2}
\end{align}
\end{widetext}
where $\kappa_\pm=1/K_w\pm 1/K_L$ and $L_\omega = v/\omega$. Taking into account boundary condition at $x=0$, i.e. $\partial_x \phi(x=0,t)=0$, we define new Green's function ${\cal G}_\omega(x,x')\rightarrow \frac{1}{2}\left[{\cal G}_\omega(x,x')+{\cal G}_\omega(-x,x')\right]$ which satisfies above boundary condition. Finally, by setting $x, x' \rightarrow 0$ one arrives at Eq.\eqref{eq:GreensL}.

\section{Boundary correlations Green's functions for the inhomogeneous Luttinger liquid model.}

We now compute the relevant correlation functions in the presence of the leads. In general, one has to compute these functions numerically. One can obtain, however, some limiting expressions for $|\delta K| \ll 1$, where $\delta K \equiv K_w-K_L$.
We first compute the correlation function ${\tilde W}(t_1-t_2)$
\begin{align}\label{eq:W_corr_leads1}
&\tilde{W}(t_1-t_2)=\frac{1}{2}\langle \left[\Phi_{cl}(t_1)-\Phi_{cl}(t_2)\right]^2\rangle\\
&=\int_{-\infty}^{\infty}\frac{d \omega}{2\pi}\left(1-\cos(\omega [t_1-t_2])\right)2i {\cal \tilde G}_{\Phi}^{K}(\omega)\nonumber\\
&=\int_{-\infty}^{\infty}\frac{d \omega}{2\pi}\left(1-\cos(\omega [t_1-t_2])\right)\frac{2\pi K_w}{\omega}\coth\left(\frac{\omega}{2T}\right)\nonumber\\
&\times\left(\frac{2K_L K_w}{K_L^2+K_w^2-(K_L^2-K_w^2)\cos(\frac{L\omega}{v})}\right)\nonumber\\ \nonumber
\end{align}
By expanding ${\cal \tilde G}_{\Phi}^{K}(\omega)$ to first order in $\delta K$, the correlation function $\tilde W(\Delta t)$ becomes
\begin{align}\label{eq:W_corr_leads2}
\!&\!{\tilde W}(t)\!\approx\! 2\!\int_{-\infty}^{\infty}\!\frac{d \omega}{\omega}\!\left(1\!-\!\cos\omega t\right)\!\left[K_w\!+\!\delta K \!\cos \frac{L\omega}{v}\!\right]\!\!\coth \frac{\omega}{2T}\nonumber\\
\!&\!=\!2K_w Y(t)\!-\!\delta K\left[\!Y\!\left(t\!+\!\frac{L}{v}\right)\!+\!Y\!\left(t\!-\!\frac{L}{v}\!\right)\!-\!2Y\!\left(\frac{L}{v}\right)\!\right]
\end{align}
where $Y(\Delta t )=\log\left(\frac{\Lambda^2}{\pi^2T^2}\!\sinh^2[\pi T\Delta t]\right)$. Here we implicitly assumed that $|\Delta t| \gg \Lambda^{-1}$.

The correlation functions $\tilde{\kappa_1}$ and $\tilde{\kappa_2}$ in the presence of the leads are given by
\begin{align}\label{eq:kappa2leads}
\!&\!i\tilde{\kappa}_1(\Delta t)\!\equiv\!\langle \delta \Phi^{(-)}_{cl}\delta \Phi^{(-)}_{q}\rangle\!=\!4i\!\int_{-\infty}^{\infty}\!\!\frac{d\omega}{2\pi}\!\left[1\!-\!\cos(\omega \Delta t)\right]{\cal \tilde G}_{\Phi}^{R}(\omega)\nonumber\\
\!&\!=2\pi i K_w \!-\!\pi i\left(2\!-\!{\rm sgn}\! \left[\frac{L}{v}\!-\!\Delta t\right]\!-\!{\rm sgn}\! \left[\frac{L}{v}\!+\!\Delta t\right]\right)\delta K\\
\!&i\tilde{\kappa}_2(\Delta t)\!\equiv\!\langle \delta \Phi^{(-)}_{cl}\delta \Phi^{(+)}_{q}\rangle=4\int_{-\infty}^{\infty}\frac{d\omega}{2\pi}\sin(\omega \Delta t){\cal \tilde G}_{\Phi}^{R}(\omega)\nonumber\\
&=-2\pi i K_w {\rm sgn} (\Delta t)\!-\! \pi i\delta K\!\left({\rm sgn}\!\left[\!\frac{L}{v}\!-\!\Delta t\!\right]\!-\!{\rm sgn} \!\left[\!\frac{L}{v}\!+\!\Delta t\!\right]\!\right).
\end{align}
Here we included only first order corrections in the small parameter $\delta K$.


\begin{thebibliography}{73}%
\makeatletter
\providecommand \@ifxundefined [1]{%
 \@ifx{#1\undefined}
}%
\providecommand \@ifnum [1]{%
 \ifnum #1\expandafter \@firstoftwo
 \else \expandafter \@secondoftwo
 \fi
}%
\providecommand \@ifx [1]{%
 \ifx #1\expandafter \@firstoftwo
 \else \expandafter \@secondoftwo
 \fi
}%
\providecommand \natexlab [1]{#1}%
\providecommand \enquote  [1]{``#1''}%
\providecommand \bibnamefont  [1]{#1}%
\providecommand \bibfnamefont [1]{#1}%
\providecommand \citenamefont [1]{#1}%
\providecommand \href@noop [0]{\@secondoftwo}%
\providecommand \href [0]{\begingroup \@sanitize@url \@href}%
\providecommand \@href[1]{\@@startlink{#1}\@@href}%
\providecommand \@@href[1]{\endgroup#1\@@endlink}%
\providecommand \@sanitize@url [0]{\catcode `\\12\catcode `\$12\catcode
  `\&12\catcode `\#12\catcode `\^12\catcode `\_12\catcode `\%12\relax}%
\providecommand \@@startlink[1]{}%
\providecommand \@@endlink[0]{}%
\providecommand \url  [0]{\begingroup\@sanitize@url \@url }%
\providecommand \@url [1]{\endgroup\@href {#1}{\urlprefix }}%
\providecommand \urlprefix  [0]{URL }%
\providecommand \Eprint [0]{\href }%
\providecommand \doibase [0]{http://dx.doi.org/}%
\providecommand \selectlanguage [0]{\@gobble}%
\providecommand \bibinfo  [0]{\@secondoftwo}%
\providecommand \bibfield  [0]{\@secondoftwo}%
\providecommand \translation [1]{[#1]}%
\providecommand \BibitemOpen [0]{}%
\providecommand \bibitemStop [0]{}%
\providecommand \bibitemNoStop [0]{.\EOS\space}%
\providecommand \EOS [0]{\spacefactor3000\relax}%
\providecommand \BibitemShut  [1]{\csname bibitem#1\endcsname}%
\let\auto@bib@innerbib\@empty
\bibitem [{\citenamefont {{Mourik}}\ \emph {et~al.}(2012)\citenamefont
  {{Mourik}}, \citenamefont {{Zuo}}, \citenamefont {{Frolov}}, \citenamefont
  {{Plissard}}, \citenamefont {{Bakkers}},\ and\ \citenamefont
  {{Kouwenhoven}}}]{Mourik2012}%
  \BibitemOpen
  \bibfield  {author} {\bibinfo {author} {\bibfnamefont {V.}~\bibnamefont
  {{Mourik}}}, \bibinfo {author} {\bibfnamefont {K.}~\bibnamefont {{Zuo}}},
  \bibinfo {author} {\bibfnamefont {S.~M.}\ \bibnamefont {{Frolov}}}, \bibinfo
  {author} {\bibfnamefont {S.~R.}\ \bibnamefont {{Plissard}}}, \bibinfo
  {author} {\bibfnamefont {E.~P.~A.~M.}\ \bibnamefont {{Bakkers}}}, \ and\
  \bibinfo {author} {\bibfnamefont {L.~P.}\ \bibnamefont {{Kouwenhoven}}},\
  }\href {\doibase 10.1126/science.1222360} {\bibfield  {journal} {\bibinfo
  {journal} {Science}\ }\textbf {\bibinfo {volume} {336}},\ \bibinfo {pages}
  {1003} (\bibinfo {year} {2012})},\ \Eprint {http://arxiv.org/abs/1204.2792}
  {arXiv:1204.2792 [cond-mat.mes-hall]} \BibitemShut {NoStop}%
\bibitem [{\citenamefont {Rokhinson}\ \emph {et~al.}()\citenamefont
  {Rokhinson}, \citenamefont {Liu},\ and\ \citenamefont
  {Furdyna}}]{Rokhinson2012}%
  \BibitemOpen
  \bibfield  {author} {\bibinfo {author} {\bibfnamefont {L.~P.}\ \bibnamefont
  {Rokhinson}}, \bibinfo {author} {\bibfnamefont {X.}~\bibnamefont {Liu}}, \
  and\ \bibinfo {author} {\bibfnamefont {J.~K.}\ \bibnamefont {Furdyna}},\
  }\href@noop {} {}\bibinfo {howpublished} {Nature Physics 8, 795
  (2012).}\BibitemShut {Stop}%
\bibitem [{\citenamefont {{Das}}\ \emph {et~al.}(2012)\citenamefont {{Das}},
  \citenamefont {{Ronen}}, \citenamefont {{Most}}, \citenamefont {{Oreg}},
  \citenamefont {{Heiblum}},\ and\ \citenamefont {{Shtrikman}}}]{Das2012}%
  \BibitemOpen
  \bibfield  {author} {\bibinfo {author} {\bibfnamefont {A.}~\bibnamefont
  {{Das}}}, \bibinfo {author} {\bibfnamefont {Y.}~\bibnamefont {{Ronen}}},
  \bibinfo {author} {\bibfnamefont {Y.}~\bibnamefont {{Most}}}, \bibinfo
  {author} {\bibfnamefont {Y.}~\bibnamefont {{Oreg}}}, \bibinfo {author}
  {\bibfnamefont {M.}~\bibnamefont {{Heiblum}}}, \ and\ \bibinfo {author}
  {\bibfnamefont {H.}~\bibnamefont {{Shtrikman}}},\ }\href {\doibase
  10.1038/nphys2479} {\bibfield  {journal} {\bibinfo  {journal} {Nature
  Physics}\ }\textbf {\bibinfo {volume} {8}},\ \bibinfo {pages} {887} (\bibinfo
  {year} {2012})},\ \Eprint {http://arxiv.org/abs/1205.7073} {arXiv:1205.7073
  [cond-mat.mes-hall]} \BibitemShut {NoStop}%
\bibitem [{\citenamefont {{Deng}}\ \emph {et~al.}(2012)\citenamefont {{Deng}},
  \citenamefont {{Yu}}, \citenamefont {{Huang}}, \citenamefont {{Larsson}},
  \citenamefont {{Caroff}},\ and\ \citenamefont {{Xu}}}]{Deng2012}%
  \BibitemOpen
  \bibfield  {author} {\bibinfo {author} {\bibfnamefont {M.~T.}\ \bibnamefont
  {{Deng}}}, \bibinfo {author} {\bibfnamefont {C.~L.}\ \bibnamefont {{Yu}}},
  \bibinfo {author} {\bibfnamefont {G.~Y.}\ \bibnamefont {{Huang}}}, \bibinfo
  {author} {\bibfnamefont {M.}~\bibnamefont {{Larsson}}}, \bibinfo {author}
  {\bibfnamefont {P.}~\bibnamefont {{Caroff}}}, \ and\ \bibinfo {author}
  {\bibfnamefont {H.~Q.}\ \bibnamefont {{Xu}}},\ }\href@noop {} {\bibfield
  {journal} {\bibinfo  {journal} {ArXiv e-prints}\ } (\bibinfo {year}
  {2012})},\ \Eprint {http://arxiv.org/abs/1204.4130} {arXiv:1204.4130
  [cond-mat.mes-hall]} \BibitemShut {NoStop}%
\bibitem [{\citenamefont {{Finck}}\ \emph {et~al.}(2012)\citenamefont
  {{Finck}}, \citenamefont {{Van Harlingen}}, \citenamefont {{Mohseni}},
  \citenamefont {{Jung}},\ and\ \citenamefont {{Li}}}]{Fink2012}%
  \BibitemOpen
  \bibfield  {author} {\bibinfo {author} {\bibfnamefont {A.~D.~K.}\
  \bibnamefont {{Finck}}}, \bibinfo {author} {\bibfnamefont {D.~J.}\
  \bibnamefont {{Van Harlingen}}}, \bibinfo {author} {\bibfnamefont {P.~K.}\
  \bibnamefont {{Mohseni}}}, \bibinfo {author} {\bibfnamefont {K.}~\bibnamefont
  {{Jung}}}, \ and\ \bibinfo {author} {\bibfnamefont {X.}~\bibnamefont
  {{Li}}},\ }\href@noop {} {\bibfield  {journal} {\bibinfo  {journal} {ArXiv
  e-prints}\ } (\bibinfo {year} {2012})},\ \Eprint
  {http://arxiv.org/abs/1212.1101} {arXiv:1212.1101 [cond-mat.mes-hall]}
  \BibitemShut {NoStop}%
\bibitem [{\citenamefont {Reich}(2012)}]{Reich}%
  \BibitemOpen
  \bibfield  {author} {\bibinfo {author} {\bibfnamefont {E.~S.}\ \bibnamefont
  {Reich}},\ }\href@noop {} {\bibfield  {journal} {\bibinfo  {journal}
  {Nature}\ }\textbf {\bibinfo {volume} {483}},\ \bibinfo {pages} {132}
  (\bibinfo {year} {2012})}\BibitemShut {NoStop}%
\bibitem [{\citenamefont {{Brouwer}}(2012)}]{Brouwer_Science}%
  \BibitemOpen
  \bibfield  {author} {\bibinfo {author} {\bibfnamefont {P.~W.}\ \bibnamefont
  {{Brouwer}}},\ }\href {\doibase 10.1126/science.1223302} {\bibfield
  {journal} {\bibinfo  {journal} {Science}\ }\textbf {\bibinfo {volume}
  {336}},\ \bibinfo {pages} {989} (\bibinfo {year} {2012})}\BibitemShut
  {NoStop}%
\bibitem [{\citenamefont {Wilczek}(2012)}]{Wilczek2012}%
  \BibitemOpen
  \bibfield  {author} {\bibinfo {author} {\bibfnamefont {F.}~\bibnamefont
  {Wilczek}},\ }\href@noop {} {\bibfield  {journal} {\bibinfo  {journal}
  {Nature}\ }\textbf {\bibinfo {volume} {486}},\ \bibinfo {pages} {195}
  (\bibinfo {year} {2012})}\BibitemShut {NoStop}%
\bibitem [{\citenamefont {{Moore}}\ and\ \citenamefont
  {{Read}}(1991)}]{Moore1991}%
  \BibitemOpen
  \bibfield  {author} {\bibinfo {author} {\bibfnamefont {G.}~\bibnamefont
  {{Moore}}}\ and\ \bibinfo {author} {\bibfnamefont {N.}~\bibnamefont
  {{Read}}},\ }\href {\doibase 10.1016/0550-3213(91)90407-O} {\bibfield
  {journal} {\bibinfo  {journal} {Nuclear Physics B}\ }\textbf {\bibinfo
  {volume} {360}},\ \bibinfo {pages} {362} (\bibinfo {year}
  {1991})}\BibitemShut {NoStop}%
\bibitem [{\citenamefont {{Nayak}}\ and\ \citenamefont
  {{Wilczek}}(1996)}]{Nayak1996}%
  \BibitemOpen
  \bibfield  {author} {\bibinfo {author} {\bibfnamefont {C.}~\bibnamefont
  {{Nayak}}}\ and\ \bibinfo {author} {\bibfnamefont {F.}~\bibnamefont
  {{Wilczek}}},\ }\href {\doibase 10.1016/0550-3213(96)00430-0} {\bibfield
  {journal} {\bibinfo  {journal} {Nuclear Physics B}\ }\textbf {\bibinfo
  {volume} {479}},\ \bibinfo {pages} {529} (\bibinfo {year} {1996})},\ \Eprint
  {http://arxiv.org/abs/arXiv:cond-mat/9605145} {arXiv:cond-mat/9605145}
  \BibitemShut {NoStop}%
\bibitem [{\citenamefont {Read}\ and\ \citenamefont {Green}(2000)}]{ReadGreen}%
  \BibitemOpen
  \bibfield  {author} {\bibinfo {author} {\bibfnamefont {N.}~\bibnamefont
  {Read}}\ and\ \bibinfo {author} {\bibfnamefont {D.}~\bibnamefont {Green}},\
  }\href@noop {} {\bibfield  {journal} {\bibinfo  {journal} {Phys.\ Rev.\ B}\
  }\textbf {\bibinfo {volume} {61}},\ \bibinfo {pages} {10267} (\bibinfo {year}
  {2000})}\BibitemShut {NoStop}%
\bibitem [{\citenamefont {Ivanov}(2001)}]{ivanov}%
  \BibitemOpen
  \bibfield  {author} {\bibinfo {author} {\bibfnamefont {D.~A.}\ \bibnamefont
  {Ivanov}},\ }\href@noop {} {\bibfield  {journal} {\bibinfo  {journal} {Phys.\
  Rev.\ Lett.}\ }\textbf {\bibinfo {volume} {86}},\ \bibinfo {pages} {268}
  (\bibinfo {year} {2001})}\BibitemShut {NoStop}%
\bibitem [{\citenamefont {Bonderson}\ \emph {et~al.}(shed)\citenamefont
  {Bonderson}, \citenamefont {Gurarie},\ and\ \citenamefont
  {Nayak}}]{BondersonNonAbelianStatistics}%
  \BibitemOpen
  \bibfield  {author} {\bibinfo {author} {\bibfnamefont {P.}~\bibnamefont
  {Bonderson}}, \bibinfo {author} {\bibfnamefont {V.}~\bibnamefont {Gurarie}},
  \ and\ \bibinfo {author} {\bibfnamefont {C.}~\bibnamefont {Nayak}},\
  }\href@noop {} {\bibfield  {journal} {\bibinfo  {journal} {arXiv:1008.5194}\
  } (\bibinfo {year} {unpublished})}\BibitemShut {NoStop}%
\bibitem [{\citenamefont {Kitaev}(2002)}]{kitaev}%
  \BibitemOpen
  \bibfield  {author} {\bibinfo {author} {\bibfnamefont {A.}~\bibnamefont
  {Kitaev}},\ }\href@noop {} {\bibfield  {journal} {\bibinfo  {journal} {Ann.\
  Phys.}\ }\textbf {\bibinfo {volume} {303}},\ \bibinfo {pages} {2} (\bibinfo
  {year} {2002})}\BibitemShut {NoStop}%
\bibitem [{\citenamefont {Nayak}\ \emph {et~al.}(2008)\citenamefont {Nayak},
  \citenamefont {Simon}, \citenamefont {Stern}, \citenamefont {Freedman},\ and\
  \citenamefont {{Das Sarma}}}]{TQCreview}%
  \BibitemOpen
  \bibfield  {author} {\bibinfo {author} {\bibfnamefont {C.}~\bibnamefont
  {Nayak}}, \bibinfo {author} {\bibfnamefont {S.~H.}\ \bibnamefont {Simon}},
  \bibinfo {author} {\bibfnamefont {A.}~\bibnamefont {Stern}}, \bibinfo
  {author} {\bibfnamefont {M.}~\bibnamefont {Freedman}}, \ and\ \bibinfo
  {author} {\bibfnamefont {S.}~\bibnamefont {{Das Sarma}}},\ }\href@noop {}
  {\bibfield  {journal} {\bibinfo  {journal} {Rev.\ Mod.\ Phys.}\ }\textbf
  {\bibinfo {volume} {80}},\ \bibinfo {pages} {1083} (\bibinfo {year}
  {2008})}\BibitemShut {NoStop}%
\bibitem [{\citenamefont {Kitaev}(2001)}]{Kitaev:2001}%
  \BibitemOpen
  \bibfield  {author} {\bibinfo {author} {\bibfnamefont {A.~Y.}\ \bibnamefont
  {Kitaev}},\ }\href@noop {} {\bibfield  {journal} {\bibinfo  {journal}
  {Physics-Uspekhi}\ }\textbf {\bibinfo {volume} {44}},\ \bibinfo {pages} {131}
  (\bibinfo {year} {2001})}\BibitemShut {NoStop}%
\bibitem [{\citenamefont {Alicea}\ \emph {et~al.}(2011)\citenamefont {Alicea},
  \citenamefont {Oreg}, \citenamefont {Refael}, \citenamefont {von Oppen},\
  and\ \citenamefont {Fisher}}]{AliceaBraiding}%
  \BibitemOpen
  \bibfield  {author} {\bibinfo {author} {\bibfnamefont {J.}~\bibnamefont
  {Alicea}}, \bibinfo {author} {\bibfnamefont {Y.}~\bibnamefont {Oreg}},
  \bibinfo {author} {\bibfnamefont {G.}~\bibnamefont {Refael}}, \bibinfo
  {author} {\bibfnamefont {F.}~\bibnamefont {von Oppen}}, \ and\ \bibinfo
  {author} {\bibfnamefont {M.~P.~A.}\ \bibnamefont {Fisher}},\ }\href@noop {}
  {\bibfield  {journal} {\bibinfo  {journal} {Nature Physics}\ }\textbf
  {\bibinfo {volume} {7}},\ \bibinfo {pages} {412} (\bibinfo {year}
  {2011})}\BibitemShut {NoStop}%
\bibitem [{\citenamefont {Sau}\ \emph {et~al.}(2010{\natexlab{a}})\citenamefont
  {Sau}, \citenamefont {Tewari},\ and\ \citenamefont {{Das
  Sarma}}}]{SauWireNetwork}%
  \BibitemOpen
  \bibfield  {author} {\bibinfo {author} {\bibfnamefont {J.~D.}\ \bibnamefont
  {Sau}}, \bibinfo {author} {\bibfnamefont {S.}~\bibnamefont {Tewari}}, \ and\
  \bibinfo {author} {\bibfnamefont {S.}~\bibnamefont {{Das Sarma}}},\ }\href
  {\doibase 10.1103/PhysRevA.82.052322} {\bibfield  {journal} {\bibinfo
  {journal} {Phys. Rev. A}\ }\textbf {\bibinfo {volume} {82}},\ \bibinfo
  {pages} {052322} (\bibinfo {year} {2010}{\natexlab{a}})}\BibitemShut
  {NoStop}%
\bibitem [{\citenamefont {Clarke}\ \emph {et~al.}(2011)\citenamefont {Clarke},
  \citenamefont {Sau},\ and\ \citenamefont {Tewari}}]{ClarkeBraiding}%
  \BibitemOpen
  \bibfield  {author} {\bibinfo {author} {\bibfnamefont {D.~J.}\ \bibnamefont
  {Clarke}}, \bibinfo {author} {\bibfnamefont {J.~D.}\ \bibnamefont {Sau}}, \
  and\ \bibinfo {author} {\bibfnamefont {S.}~\bibnamefont {Tewari}},\ }\href
  {\doibase 10.1103/PhysRevB.84.035120} {\bibfield  {journal} {\bibinfo
  {journal} {Phys. Rev. B}\ }\textbf {\bibinfo {volume} {84}},\ \bibinfo
  {pages} {035120} (\bibinfo {year} {2011})}\BibitemShut {NoStop}%
\bibitem [{\citenamefont {Bonderson}\ and\ \citenamefont
  {Lutchyn}(2011)}]{TopologicalQuantumBus}%
  \BibitemOpen
  \bibfield  {author} {\bibinfo {author} {\bibfnamefont {P.}~\bibnamefont
  {Bonderson}}\ and\ \bibinfo {author} {\bibfnamefont {R.~M.}\ \bibnamefont
  {Lutchyn}},\ }\href {\doibase 10.1103/PhysRevLett.106.130505} {\bibfield
  {journal} {\bibinfo  {journal} {Phys. Rev. Lett.}\ }\textbf {\bibinfo
  {volume} {106}},\ \bibinfo {pages} {130505} (\bibinfo {year}
  {2011})}\BibitemShut {NoStop}%
\bibitem [{\citenamefont {Fu}\ and\ \citenamefont {Kane}(2008)}]{Fu:2008}%
  \BibitemOpen
  \bibfield  {author} {\bibinfo {author} {\bibfnamefont {L.}~\bibnamefont
  {Fu}}\ and\ \bibinfo {author} {\bibfnamefont {C. L.}~\bibnamefont {Kane}},\
  }\href@noop {} {\bibfield  {journal} {\bibinfo  {journal} {Phys. Rev. Lett.}\
  }\textbf {\bibinfo {volume} {100}},\ \bibinfo {pages} {096407} (\bibinfo
  {year} {2008})}\BibitemShut {NoStop}%
\bibitem [{\citenamefont {Fu}\ and\ \citenamefont
  {Kane}(2009)}]{MajoranaQSHedge}%
  \BibitemOpen
  \bibfield  {author} {\bibinfo {author} {\bibfnamefont {L.}~\bibnamefont
  {Fu}}\ and\ \bibinfo {author} {\bibfnamefont {C.~L.}\ \bibnamefont {Kane}},\
  }\href@noop {} {\bibfield  {journal} {\bibinfo  {journal} {Phys.\ Rev.\ B}\
  }\textbf {\bibinfo {volume} {79}},\ \bibinfo {pages} {161408(R)} (\bibinfo
  {year} {2009})}\BibitemShut {NoStop}%
\bibitem [{\citenamefont {Sau}\ \emph {et~al.}(2010{\natexlab{b}})\citenamefont
  {Sau}, \citenamefont {Lutchyn}, \citenamefont {Tewari},\ and\ \citenamefont
  {{Das Sarma}}}]{Sau}%
  \BibitemOpen
  \bibfield  {author} {\bibinfo {author} {\bibfnamefont {J.~D.}\ \bibnamefont
  {Sau}}, \bibinfo {author} {\bibfnamefont {R.~M.}\ \bibnamefont {Lutchyn}},
  \bibinfo {author} {\bibfnamefont {S.}~\bibnamefont {Tewari}}, \ and\ \bibinfo
  {author} {\bibfnamefont {S.}~\bibnamefont {{Das Sarma}}},\ }\href@noop {}
  {\bibfield  {journal} {\bibinfo  {journal} {Phys.\ Rev.\ Lett.}\ }\textbf
  {\bibinfo {volume} {104}},\ \bibinfo {pages} {040502} (\bibinfo {year}
  {2010}{\natexlab{b}})}\BibitemShut {NoStop}%
\bibitem [{\citenamefont {Alicea}(2010)}]{Alicea}%
  \BibitemOpen
  \bibfield  {author} {\bibinfo {author} {\bibfnamefont {J.}~\bibnamefont
  {Alicea}},\ }\href@noop {} {\bibfield  {journal} {\bibinfo  {journal} {Phys.\
  Rev.\ B}\ }\textbf {\bibinfo {volume} {81}},\ \bibinfo {pages} {125318}
  (\bibinfo {year} {2010})}\BibitemShut {NoStop}%
\bibitem [{\citenamefont {Lutchyn}\ \emph {et~al.}(2010)\citenamefont
  {Lutchyn}, \citenamefont {Sau},\ and\ \citenamefont
  {Das~Sarma}}]{1DwiresLutchyn}%
  \BibitemOpen
  \bibfield  {author} {\bibinfo {author} {\bibfnamefont {R.~M.}\ \bibnamefont
  {Lutchyn}}, \bibinfo {author} {\bibfnamefont {J.~D.}\ \bibnamefont {Sau}}, \
  and\ \bibinfo {author} {\bibfnamefont {S.}~\bibnamefont {Das~Sarma}},\
  }\href@noop {} {\bibfield  {journal} {\bibinfo  {journal} {Phys.\ Rev.\
  Lett.}\ }\textbf {\bibinfo {volume} {105}},\ \bibinfo {pages} {077001}
  (\bibinfo {year} {2010})}\BibitemShut {NoStop}%
\bibitem [{\citenamefont {Oreg}\ \emph {et~al.}(2010)\citenamefont {Oreg},
  \citenamefont {Refael},\ and\ \citenamefont {von Oppen}}]{1DwiresOreg}%
  \BibitemOpen
  \bibfield  {author} {\bibinfo {author} {\bibfnamefont {Y.}~\bibnamefont
  {Oreg}}, \bibinfo {author} {\bibfnamefont {G.}~\bibnamefont {Refael}}, \ and\
  \bibinfo {author} {\bibfnamefont {F.}~\bibnamefont {von Oppen}},\ }\href@noop
  {} {\bibfield  {journal} {\bibinfo  {journal} {Phys.\ Rev.\ Lett.}\ }\textbf
  {\bibinfo {volume} {105}},\ \bibinfo {pages} {177002} (\bibinfo {year}
  {2010})}\BibitemShut {NoStop}%
\bibitem [{\citenamefont {Beenakker}(shed)}]{BeenakkerReview}%
  \BibitemOpen
  \bibfield  {author} {\bibinfo {author} {\bibfnamefont {C.~W.~J.}\
  \bibnamefont {Beenakker}},\ }\href@noop {} {\bibfield  {journal} {\bibinfo
  {journal} {arXiv:1112.1950}\ } (\bibinfo {year} {unpublished})}\BibitemShut
  {NoStop}%
\bibitem [{\citenamefont {Alicea}(shed)}]{AliceaReview}%
  \BibitemOpen
  \bibfield  {author} {\bibinfo {author} {\bibfnamefont {J.}~\bibnamefont
  {Alicea}},\ }\href@noop {} {\bibfield  {journal} {\bibinfo  {journal}
  {arXiv:1202.1293}\ } (\bibinfo {year} {unpublished})}\BibitemShut {NoStop}%
\bibitem [{\citenamefont {Sengupta}\ \emph {et~al.}(2001)\citenamefont
  {Sengupta}, \citenamefont {\ifmmode \check{Z}\else
  \v{Z}\fi{}uti\ifmmode~\acute{c}\else \'{c}\fi{}}, \citenamefont {Kwon},
  \citenamefont {Yakovenko},\ and\ \citenamefont
  {Das~Sarma}}]{ZeroBiasAnomaly0}%
  \BibitemOpen
  \bibfield  {author} {\bibinfo {author} {\bibfnamefont {K.}~\bibnamefont
  {Sengupta}}, \bibinfo {author} {\bibfnamefont {I.}~\bibnamefont {Zutic}}, \bibinfo
  {author} {\bibfnamefont {H.-J.}\ \bibnamefont {Kwon}}, \bibinfo {author}
  {\bibfnamefont {V.~M.}\ \bibnamefont {Yakovenko}}, \ and\ \bibinfo {author}
  {\bibfnamefont {S.}~\bibnamefont {Das~Sarma}},\ }\href {\doibase
  10.1103/PhysRevB.63.144531} {\bibfield  {journal} {\bibinfo  {journal} {Phys.
  Rev. B}\ }\textbf {\bibinfo {volume} {63}},\ \bibinfo {pages} {144531}
  (\bibinfo {year} {2001})}\BibitemShut {NoStop}%
\bibitem [{\citenamefont {Bolech}\ and\ \citenamefont
  {Demler}(2007)}]{ZeroBiasAnomaly1}%
  \BibitemOpen
  \bibfield  {author} {\bibinfo {author} {\bibfnamefont {C.~J.}\ \bibnamefont
  {Bolech}}\ and\ \bibinfo {author} {\bibfnamefont {E.}~\bibnamefont
  {Demler}},\ }\href {\doibase 10.1103/PhysRevLett.98.237002} {\bibfield
  {journal} {\bibinfo  {journal} {Phys. Rev. Lett.}\ }\textbf {\bibinfo
  {volume} {98}},\ \bibinfo {pages} {237002} (\bibinfo {year}
  {2007})}\BibitemShut {NoStop}%
\bibitem [{\citenamefont {Nilsson}\ \emph {et~al.}(2008)\citenamefont
  {Nilsson}, \citenamefont {Akhmerov},\ and\ \citenamefont
  {Beenakker}}]{ZeroBiasAnomaly2}%
  \BibitemOpen
  \bibfield  {author} {\bibinfo {author} {\bibfnamefont {J.}~\bibnamefont
  {Nilsson}}, \bibinfo {author} {\bibfnamefont {A.~R.}\ \bibnamefont
  {Akhmerov}}, \ and\ \bibinfo {author} {\bibfnamefont {C.~W.~J.}\ \bibnamefont
  {Beenakker}},\ }\href {\doibase 10.1103/PhysRevLett.101.120403} {\bibfield
  {journal} {\bibinfo  {journal} {Phys. Rev. Lett.}\ }\textbf {\bibinfo
  {volume} {101}},\ \bibinfo {pages} {120403} (\bibinfo {year}
  {2008})}\BibitemShut {NoStop}%
\bibitem [{\citenamefont {Law}\ \emph {et~al.}(2009)\citenamefont {Law},
  \citenamefont {Lee},\ and\ \citenamefont {Ng}}]{ZeroBiasAnomaly3}%
  \BibitemOpen
  \bibfield  {author} {\bibinfo {author} {\bibfnamefont {K.~T.}\ \bibnamefont
  {Law}}, \bibinfo {author} {\bibfnamefont {P.~A.}\ \bibnamefont {Lee}}, \ and\
  \bibinfo {author} {\bibfnamefont {T.~K.}\ \bibnamefont {Ng}},\ }\href@noop {}
  {\bibfield  {journal} {\bibinfo  {journal} {Phys.\ Rev.\ Lett.}\ }\textbf
  {\bibinfo {volume} {103}},\ \bibinfo {pages} {237001} (\bibinfo {year}
  {2009})}\BibitemShut {NoStop}%
\bibitem [{\citenamefont {Sau}\ \emph {et~al.}(2010{\natexlab{c}})\citenamefont
  {Sau}, \citenamefont {Tewari}, \citenamefont {Lutchyn}, \citenamefont
  {Stanescu},\ and\ \citenamefont {Das~Sarma}}]{ZeroBiasAnomaly31}%
  \BibitemOpen
  \bibfield  {author} {\bibinfo {author} {\bibfnamefont {J.~D.}\ \bibnamefont
  {Sau}}, \bibinfo {author} {\bibfnamefont {S.}~\bibnamefont {Tewari}},
  \bibinfo {author} {\bibfnamefont {R.~M.}\ \bibnamefont {Lutchyn}}, \bibinfo
  {author} {\bibfnamefont {T.~D.}\ \bibnamefont {Stanescu}}, \ and\ \bibinfo
  {author} {\bibfnamefont {S.}~\bibnamefont {Das~Sarma}},\ }\href {\doibase
  10.1103/PhysRevB.82.214509} {\bibfield  {journal} {\bibinfo  {journal} {Phys.
  Rev. B}\ }\textbf {\bibinfo {volume} {82}},\ \bibinfo {pages} {214509}
  (\bibinfo {year} {2010}{\natexlab{c}})}\BibitemShut {NoStop}%
\bibitem [{\citenamefont {Flensberg}(2010)}]{ZeroBiasAnomaly4}%
  \BibitemOpen
  \bibfield  {author} {\bibinfo {author} {\bibfnamefont {K.}~\bibnamefont
  {Flensberg}},\ }\href {\doibase 10.1103/PhysRevB.82.180516} {\bibfield
  {journal} {\bibinfo  {journal} {Phys. Rev. B}\ }\textbf {\bibinfo {volume}
  {82}},\ \bibinfo {pages} {180516} (\bibinfo {year} {2010})}\BibitemShut
  {NoStop}%
\bibitem [{\citenamefont {Golub}\ and\ \citenamefont
  {Horovitz}(2011)}]{ZeroBiasAnomaly5}%
  \BibitemOpen
  \bibfield  {author} {\bibinfo {author} {\bibfnamefont {A.}~\bibnamefont
  {Golub}}\ and\ \bibinfo {author} {\bibfnamefont {B.}~\bibnamefont
  {Horovitz}},\ }\href {\doibase 10.1103/PhysRevB.83.153415} {\bibfield
  {journal} {\bibinfo  {journal} {Phys. Rev. B}\ }\textbf {\bibinfo {volume}
  {83}},\ \bibinfo {pages} {153415} (\bibinfo {year} {2011})}\BibitemShut
  {NoStop}%
\bibitem [{\citenamefont {Wimmer}\ \emph {et~al.}(2011)\citenamefont {Wimmer},
  \citenamefont {Akhmerov}, \citenamefont {Dahlhaus},\ and\ \citenamefont
  {Beenakker}}]{ZeroBiasAnomaly6}%
  \BibitemOpen
  \bibfield  {author} {\bibinfo {author} {\bibfnamefont {M.}~\bibnamefont
  {Wimmer}}, \bibinfo {author} {\bibfnamefont {A.~R.}\ \bibnamefont
  {Akhmerov}}, \bibinfo {author} {\bibfnamefont {J.~P.}\ \bibnamefont
  {Dahlhaus}}, \ and\ \bibinfo {author} {\bibfnamefont {C.~W.~J.}\ \bibnamefont
  {Beenakker}},\ }\href@noop {} {\bibfield  {journal} {\bibinfo  {journal} {New
  Journal of Physics}\ }\textbf {\bibinfo {volume} {13}},\ \bibinfo {pages}
  {053016} (\bibinfo {year} {2011})}\BibitemShut {NoStop}%
\bibitem [{\citenamefont {Stanescu}\ \emph {et~al.}(2011)\citenamefont
  {Stanescu}, \citenamefont {Lutchyn},\ and\ \citenamefont
  {Das~Sarma}}]{ZeroBiasAnomaly61}%
  \BibitemOpen
  \bibfield  {author} {\bibinfo {author} {\bibfnamefont {T.~D.}\ \bibnamefont
  {Stanescu}}, \bibinfo {author} {\bibfnamefont {R.~M.}\ \bibnamefont
  {Lutchyn}}, \ and\ \bibinfo {author} {\bibfnamefont {S.}~\bibnamefont
  {Das~Sarma}},\ }\href {\doibase 10.1103/PhysRevB.84.144522} {\bibfield
  {journal} {\bibinfo  {journal} {Phys. Rev. B}\ }\textbf {\bibinfo {volume}
  {84}},\ \bibinfo {pages} {144522} (\bibinfo {year} {2011})}\BibitemShut
  {NoStop}%
\bibitem [{\citenamefont {Qu}\ \emph {et~al.}(shed)\citenamefont {Qu},
  \citenamefont {Zhang}, \citenamefont {Mao},\ and\ \citenamefont
  {Zhang}}]{ZeroBiasAnomaly7}%
  \BibitemOpen
  \bibfield  {author} {\bibinfo {author} {\bibfnamefont {C.}~\bibnamefont
  {Qu}}, \bibinfo {author} {\bibfnamefont {Y.}~\bibnamefont {Zhang}}, \bibinfo
  {author} {\bibfnamefont {L.}~\bibnamefont {Mao}}, \ and\ \bibinfo {author}
  {\bibfnamefont {C.}~\bibnamefont {Zhang}},\ }\href@noop {} {\bibfield
  {journal} {\bibinfo  {journal} {arXiv:1109.4108}\ } (\bibinfo {year}
  {unpublished})}\BibitemShut {NoStop}%
\bibitem [{\citenamefont {Giamarchi}(2004)}]{Giamarchi:2004}%
  \BibitemOpen
  \bibfield  {author} {\bibinfo {author} {\bibfnamefont {T.}~\bibnamefont
  {Giamarchi}},\ }\href@noop {} {\emph {\bibinfo {title} {Quantum Physics in
  One Dimension}}},\ New York\ (\bibinfo  {publisher} {Oxford University
  Press},\ \bibinfo {year} {2004})\BibitemShut {NoStop}%
\bibitem [{\citenamefont {{Fidkowski}}\ \emph {et~al.}(2012)\citenamefont
  {{Fidkowski}}, \citenamefont {{Alicea}}, \citenamefont {{Lindner}},
  \citenamefont {{Lutchyn}},\ and\ \citenamefont {{Fisher}}}]{Fidkowski2012}%
  \BibitemOpen
  \bibfield  {author} {\bibinfo {author} {\bibfnamefont {L.}~\bibnamefont
  {{Fidkowski}}}, \bibinfo {author} {\bibfnamefont {J.}~\bibnamefont
  {{Alicea}}}, \bibinfo {author} {\bibfnamefont {N.~H.}\ \bibnamefont
  {{Lindner}}}, \bibinfo {author} {\bibfnamefont {R.~M.}\ \bibnamefont
  {{Lutchyn}}}, \ and\ \bibinfo {author} {\bibfnamefont {M.~P.~A.}\
  \bibnamefont {{Fisher}}},\ }\href {\doibase 10.1103/PhysRevB.85.245121}
  {\bibfield  {journal} {\bibinfo  {journal} {\prb}\ }\textbf {\bibinfo
  {volume} {85}},\ \bibinfo {eid} {245121} (\bibinfo {year} {2012})},\ \Eprint
  {http://arxiv.org/abs/1203.4818} {arXiv:1203.4818 [cond-mat.str-el]}
  \BibitemShut {NoStop}%
\bibitem [{\citenamefont {{Maslov}}\ \emph {et~al.}(1996)\citenamefont
  {{Maslov}}, \citenamefont {{Stone}}, \citenamefont {{Goldbart}},\ and\
  \citenamefont {{Loss}}}]{maslov}%
  \BibitemOpen
  \bibfield  {author} {\bibinfo {author} {\bibfnamefont {D.~L.}\ \bibnamefont
  {{Maslov}}}, \bibinfo {author} {\bibfnamefont {M.}~\bibnamefont {{Stone}}},
  \bibinfo {author} {\bibfnamefont {P.~M.}\ \bibnamefont {{Goldbart}}}, \ and\
  \bibinfo {author} {\bibfnamefont {D.}~\bibnamefont {{Loss}}},\ }\href
  {\doibase 10.1103/PhysRevB.53.1548} {\bibfield  {journal} {\bibinfo
  {journal} {\prb}\ }\textbf {\bibinfo {volume} {53}},\ \bibinfo {pages} {1548}
  (\bibinfo {year} {1996})}\BibitemShut {NoStop}%
\bibitem [{\citenamefont {{Affleck}}\ \emph {et~al.}(2000)\citenamefont
  {{Affleck}}, \citenamefont {{Caux}},\ and\ \citenamefont
  {{Zagoskin}}}]{Affleck2000}%
  \BibitemOpen
  \bibfield  {author} {\bibinfo {author} {\bibfnamefont {I.}~\bibnamefont
  {{Affleck}}}, \bibinfo {author} {\bibfnamefont {J.-S.}\ \bibnamefont
  {{Caux}}}, \ and\ \bibinfo {author} {\bibfnamefont {A.~M.}\ \bibnamefont
  {{Zagoskin}}},\ }\href {\doibase 10.1103/PhysRevB.62.1433} {\bibfield
  {journal} {\bibinfo  {journal} {\prb}\ }\textbf {\bibinfo {volume} {62}},\
  \bibinfo {pages} {1433} (\bibinfo {year} {2000})},\ \Eprint
  {http://arxiv.org/abs/arXiv:cond-mat/0001375} {arXiv:cond-mat/0001375}
  \BibitemShut {NoStop}%
\bibitem [{\citenamefont {Winkelholz}\ \emph {et~al.}(1996)\citenamefont
  {Winkelholz}, \citenamefont {Fazio}, \citenamefont {Hekking},\ and\
  \citenamefont {Sch\"on}}]{LDOSdivergence}%
  \BibitemOpen
  \bibfield  {author} {\bibinfo {author} {\bibfnamefont {C.}~\bibnamefont
  {Winkelholz}}, \bibinfo {author} {\bibfnamefont {R.}~\bibnamefont {Fazio}},
  \bibinfo {author} {\bibfnamefont {F.~W.~J.}\ \bibnamefont {Hekking}}, \ and\
  \bibinfo {author} {\bibfnamefont {G.}~\bibnamefont {Sch\"on}},\ }\href
  {\doibase 10.1103/PhysRevLett.77.3200} {\bibfield  {journal} {\bibinfo
  {journal} {Phys. Rev. Lett.}\ }\textbf {\bibinfo {volume} {77}},\ \bibinfo
  {pages} {3200} (\bibinfo {year} {1996})}\BibitemShut {NoStop}%
\bibitem [{\citenamefont {{Apel}}\ and\ \citenamefont
  {{Rice}}(1982)}]{Apel1982}%
  \BibitemOpen
  \bibfield  {author} {\bibinfo {author} {\bibfnamefont {W.}~\bibnamefont
  {{Apel}}}\ and\ \bibinfo {author} {\bibfnamefont {T.~M.}\ \bibnamefont
  {{Rice}}},\ }\href {\doibase 10.1103/PhysRevB.26.7063} {\bibfield  {journal}
  {\bibinfo  {journal} {\prb}\ }\textbf {\bibinfo {volume} {26}},\ \bibinfo
  {pages} {7063} (\bibinfo {year} {1982})}\BibitemShut {NoStop}%
\bibitem [{\citenamefont {Kane}\ and\ \citenamefont
  {Fisher}(1992{\natexlab{a}})}]{KaneFisher1}%
  \BibitemOpen
  \bibfield  {author} {\bibinfo {author} {\bibfnamefont {C.~L.}\ \bibnamefont
  {Kane}}\ and\ \bibinfo {author} {\bibfnamefont {M.~P.~A.}\ \bibnamefont
  {Fisher}},\ }\href {\doibase 10.1103/PhysRevLett.68.1220} {\bibfield
  {journal} {\bibinfo  {journal} {Phys. Rev. Lett.}\ }\textbf {\bibinfo
  {volume} {68}},\ \bibinfo {pages} {1220} (\bibinfo {year}
  {1992}{\natexlab{a}})}\BibitemShut {NoStop}%
\bibitem [{\citenamefont {{Wen}}(1991)}]{Wen1991}%
  \BibitemOpen
  \bibfield  {author} {\bibinfo {author} {\bibfnamefont {X.-G.}\ \bibnamefont
  {{Wen}}},\ }\href {\doibase 10.1103/PhysRevB.44.5708} {\bibfield  {journal}
  {\bibinfo  {journal} {\prb}\ }\textbf {\bibinfo {volume} {44}},\ \bibinfo
  {pages} {5708} (\bibinfo {year} {1991})}\BibitemShut {NoStop}%
\bibitem [{\citenamefont {{Matveev}}\ \emph {et~al.}(1993)\citenamefont
  {{Matveev}}, \citenamefont {{Yue}},\ and\ \citenamefont
  {{Glazman}}}]{Matveev1993}%
  \BibitemOpen
  \bibfield  {author} {\bibinfo {author} {\bibfnamefont {K.~A.}\ \bibnamefont
  {{Matveev}}}, \bibinfo {author} {\bibfnamefont {D.}~\bibnamefont {{Yue}}}, \
  and\ \bibinfo {author} {\bibfnamefont {L.~I.}\ \bibnamefont {{Glazman}}},\
  }\href {\doibase 10.1103/PhysRevLett.71.3351} {\bibfield  {journal} {\bibinfo
   {journal} {Physical Review Letters}\ }\textbf {\bibinfo {volume} {71}},\
  \bibinfo {pages} {3351} (\bibinfo {year} {1993})},\ \Eprint
  {http://arxiv.org/abs/arXiv:cond-mat/9306041} {arXiv:cond-mat/9306041}
  \BibitemShut {NoStop}%
\bibitem [{\citenamefont {{Furusaki}}\ and\ \citenamefont
  {{Nagaosa}}(1993)}]{Furusaki1993}%
  \BibitemOpen
  \bibfield  {author} {\bibinfo {author} {\bibfnamefont {A.}~\bibnamefont
  {{Furusaki}}}\ and\ \bibinfo {author} {\bibfnamefont {N.}~\bibnamefont
  {{Nagaosa}}},\ }\href {\doibase 10.1103/PhysRevB.47.4631} {\bibfield
  {journal} {\bibinfo  {journal} {\prb}\ }\textbf {\bibinfo {volume} {47}},\
  \bibinfo {pages} {4631} (\bibinfo {year} {1993})}\BibitemShut {NoStop}%
\bibitem [{\citenamefont {Safi}\ and\ \citenamefont
  {Schulz}(1995)}]{Safi_Schulz}%
  \BibitemOpen
  \bibfield  {author} {\bibinfo {author} {\bibfnamefont {I.}~\bibnamefont
  {Safi}}\ and\ \bibinfo {author} {\bibfnamefont {H.~J.}\ \bibnamefont
  {Schulz}},\ }\href {\doibase 10.1103/PhysRevB.52.R17040} {\bibfield
  {journal} {\bibinfo  {journal} {Phys. Rev. B}\ }\textbf {\bibinfo {volume}
  {52}},\ \bibinfo {pages} {R17040} (\bibinfo {year} {1995})}\BibitemShut
  {NoStop}%
\bibitem [{\citenamefont {{Maslov}}\ and\ \citenamefont
  {{Stone}}(1995)}]{Maslov1995}%
  \BibitemOpen
  \bibfield  {author} {\bibinfo {author} {\bibfnamefont {D.~L.}\ \bibnamefont
  {{Maslov}}}\ and\ \bibinfo {author} {\bibfnamefont {M.}~\bibnamefont
  {{Stone}}},\ }\href {\doibase 10.1103/PhysRevB.52.R5539} {\bibfield
  {journal} {\bibinfo  {journal} {\prb}\ }\textbf {\bibinfo {volume} {52}},\
  \bibinfo {pages} {5539} (\bibinfo {year} {1995})},\ \Eprint
  {http://arxiv.org/abs/arXiv:cond-mat/9505098} {arXiv:cond-mat/9505098}
  \BibitemShut {NoStop}%
\bibitem [{\citenamefont {{Ponomarenko}}(1995)}]{Ponomarenko1995}%
  \BibitemOpen
  \bibfield  {author} {\bibinfo {author} {\bibfnamefont {V.~V.}\ \bibnamefont
  {{Ponomarenko}}},\ }\href {\doibase 10.1103/PhysRevB.52.R8666} {\bibfield
  {journal} {\bibinfo  {journal} {\prb}\ }\textbf {\bibinfo {volume} {52}},\
  \bibinfo {pages} {8666} (\bibinfo {year} {1995})},\ \Eprint
  {http://arxiv.org/abs/arXiv:cond-mat/9511024} {arXiv:cond-mat/9511024}
  \BibitemShut {NoStop}%
\bibitem [{\citenamefont {Oreg}\ and\ \citenamefont
  {Finkel'stein}(1996)}]{Oreg1995}%
  \BibitemOpen
  \bibfield  {author} {\bibinfo {author} {\bibfnamefont {Y.}~\bibnamefont
  {Oreg}}\ and\ \bibinfo {author} {\bibfnamefont {A.~M.}\ \bibnamefont
  {Finkel'stein}},\ }\href {\doibase 10.1103/PhysRevB.54.R14265} {\bibfield
  {journal} {\bibinfo  {journal} {Phys. Rev. B}\ }\textbf {\bibinfo {volume}
  {54}},\ \bibinfo {pages} {R14265} (\bibinfo {year} {1996})}\BibitemShut
  {NoStop}%
\bibitem [{\citenamefont {{Alekseev}}\ \emph {et~al.}(1996)\citenamefont
  {{Alekseev}}, \citenamefont {{Cheianov}},\ and\ \citenamefont
  {{Fr{\"o}hlich}}}]{Alekseev1996}%
  \BibitemOpen
  \bibfield  {author} {\bibinfo {author} {\bibfnamefont {A.~Y.}\ \bibnamefont
  {{Alekseev}}}, \bibinfo {author} {\bibfnamefont {V.~V.}\ \bibnamefont
  {{Cheianov}}}, \ and\ \bibinfo {author} {\bibfnamefont {J.}~\bibnamefont
  {{Fr{\"o}hlich}}},\ }\href {\doibase 10.1103/PhysRevB.54.R17320} {\bibfield
  {journal} {\bibinfo  {journal} {\prb}\ }\textbf {\bibinfo {volume} {54}},\
  \bibinfo {pages} {17320} (\bibinfo {year} {1996})},\ \Eprint
  {http://arxiv.org/abs/arXiv:cond-mat/9607144} {arXiv:cond-mat/9607144}
  \BibitemShut {NoStop}%
\bibitem [{\citenamefont {{Chamon}}\ and\ \citenamefont
  {{Fradkin}}(1996)}]{Chamon1996}%
  \BibitemOpen
  \bibfield  {author} {\bibinfo {author} {\bibfnamefont {C.~d.~C.}\
  \bibnamefont {{Chamon}}}\ and\ \bibinfo {author} {\bibfnamefont
  {E.}~\bibnamefont {{Fradkin}}},\ }\href@noop {} {\bibfield  {journal}
  {\bibinfo  {journal} {eprint arXiv:cond-mat/9612185}\ } (\bibinfo {year}
  {1996})},\ \Eprint {http://arxiv.org/abs/arXiv:cond-mat/9612185}
  {arXiv:cond-mat/9612185} \BibitemShut {NoStop}%
\bibitem [{\citenamefont {Dolcini}\ \emph {et~al.}(2003)\citenamefont
  {Dolcini}, \citenamefont {Grabert}, \citenamefont {Safi},\ and\ \citenamefont
  {Trauzettel}}]{DolciniPRL'03}%
  \BibitemOpen
  \bibfield  {author} {\bibinfo {author} {\bibfnamefont {F.}~\bibnamefont
  {Dolcini}}, \bibinfo {author} {\bibfnamefont {H.}~\bibnamefont {Grabert}},
  \bibinfo {author} {\bibfnamefont {I.}~\bibnamefont {Safi}}, \ and\ \bibinfo
  {author} {\bibfnamefont {B.}~\bibnamefont {Trauzettel}},\ }\href {\doibase
  10.1103/PhysRevLett.91.266402} {\bibfield  {journal} {\bibinfo  {journal}
  {Phys. Rev. Lett.}\ }\textbf {\bibinfo {volume} {91}},\ \bibinfo {pages}
  {266402} (\bibinfo {year} {2003})}\BibitemShut {NoStop}%
\bibitem [{\citenamefont {Dolcini}\ \emph {et~al.}(2005)\citenamefont
  {Dolcini}, \citenamefont {Trauzettel}, \citenamefont {Safi},\ and\
  \citenamefont {Grabert}}]{DolciniPRB'05}%
  \BibitemOpen
  \bibfield  {author} {\bibinfo {author} {\bibfnamefont {F.}~\bibnamefont
  {Dolcini}}, \bibinfo {author} {\bibfnamefont {B.}~\bibnamefont {Trauzettel}},
  \bibinfo {author} {\bibfnamefont {I.}~\bibnamefont {Safi}}, \ and\ \bibinfo
  {author} {\bibfnamefont {H.}~\bibnamefont {Grabert}},\ }\href {\doibase
  10.1103/PhysRevB.71.165309} {\bibfield  {journal} {\bibinfo  {journal} {Phys.
  Rev. B}\ }\textbf {\bibinfo {volume} {71}},\ \bibinfo {pages} {165309}
  (\bibinfo {year} {2005})}\BibitemShut {NoStop}%
\bibitem{Thomale} R. Thomale and A. Seidel, Phys. Rev. B 83, 115330 (2011)
\bibitem [{\citenamefont {{Affleck}}(2008)}]{Affleck2008}%
  \BibitemOpen
  \bibfield  {author} {\bibinfo {author} {\bibfnamefont {I.}~\bibnamefont
  {{Affleck}}},\ }\href@noop {} {\bibfield  {journal} {\bibinfo  {journal}
  {ArXiv e-prints}\ } (\bibinfo {year} {2008})},\ \Eprint
  {http://arxiv.org/abs/0809.3474} {arXiv:0809.3474 [cond-mat.str-el]}
  \BibitemShut {NoStop}%
\bibitem [{\citenamefont {{Fendley}}\ \emph {et~al.}(1995)\citenamefont
  {{Fendley}}, \citenamefont {{Ludwig}},\ and\ \citenamefont
  {{Saleur}}}]{Fendley1995}%
  \BibitemOpen
  \bibfield  {author} {\bibinfo {author} {\bibfnamefont {P.}~\bibnamefont
  {{Fendley}}}, \bibinfo {author} {\bibfnamefont {A.~W.~W.}\ \bibnamefont
  {{Ludwig}}}, \ and\ \bibinfo {author} {\bibfnamefont {H.}~\bibnamefont
  {{Saleur}}},\ }\href {\doibase 10.1103/PhysRevLett.74.3005} {\bibfield
  {journal} {\bibinfo  {journal} {Physical Review Letters}\ }\textbf {\bibinfo
  {volume} {74}},\ \bibinfo {pages} {3005} (\bibinfo {year} {1995})},\ \Eprint
  {http://arxiv.org/abs/arXiv:cond-mat/9408068} {arXiv:cond-mat/9408068}
  \BibitemShut {NoStop}%
\bibitem [{\citenamefont {Wimmer}\ \emph {et~al.}(2010)\citenamefont {Wimmer},
  \citenamefont {Akhmerov}, \citenamefont {Medvedyeva}, \citenamefont
  {Tworzyd\l{}o},\ and\ \citenamefont {Beenakker}}]{WimmerMultichannel}%
  \BibitemOpen
  \bibfield  {author} {\bibinfo {author} {\bibfnamefont {M.}~\bibnamefont
  {Wimmer}}, \bibinfo {author} {\bibfnamefont {A.~R.}\ \bibnamefont
  {Akhmerov}}, \bibinfo {author} {\bibfnamefont {M.~V.}\ \bibnamefont
  {Medvedyeva}}, \bibinfo {author} {\bibfnamefont {J.}~\bibnamefont
  {Tworzyd\l{}o}}, \ and\ \bibinfo {author} {\bibfnamefont {C.~W.~J.}\
  \bibnamefont {Beenakker}},\ }\href {\doibase 10.1103/PhysRevLett.105.046803}
  {\bibfield  {journal} {\bibinfo  {journal} {Phys. Rev. Lett.}\ }\textbf
  {\bibinfo {volume} {105}},\ \bibinfo {pages} {046803} (\bibinfo {year}
  {2010})}\BibitemShut {NoStop}%
\bibitem [{\citenamefont {Potter}\ and\ \citenamefont
  {Lee}(2010)}]{1DwiresPotter}%
  \BibitemOpen
  \bibfield  {author} {\bibinfo {author} {\bibfnamefont {A.~C.}\ \bibnamefont
  {Potter}}\ and\ \bibinfo {author} {\bibfnamefont {P.~A.}\ \bibnamefont
  {Lee}},\ }\href {\doibase 10.1103/PhysRevLett.105.227003} {\bibfield
  {journal} {\bibinfo  {journal} {Phys. Rev. Lett.}\ }\textbf {\bibinfo
  {volume} {105}},\ \bibinfo {pages} {227003} (\bibinfo {year}
  {2010})}\BibitemShut {NoStop}%
\bibitem [{\citenamefont {{Lutchyn}}\ \emph {et~al.}(2011)\citenamefont
  {{Lutchyn}}, \citenamefont {{Stanescu}},\ and\ \citenamefont {{Das
  Sarma}}}]{1DwiresLutchyn2}%
  \BibitemOpen
  \bibfield  {author} {\bibinfo {author} {\bibfnamefont {R.~M.}\ \bibnamefont
  {{Lutchyn}}}, \bibinfo {author} {\bibfnamefont {T.~D.}\ \bibnamefont
  {{Stanescu}}}, \ and\ \bibinfo {author} {\bibfnamefont {S.}~\bibnamefont
  {{Das Sarma}}},\ }\href {\doibase 10.1103/PhysRevLett.106.127001} {\bibfield
  {journal} {\bibinfo  {journal} {Physical Review Letters}\ }\textbf {\bibinfo
  {volume} {106}},\ \bibinfo {eid} {127001} (\bibinfo {year} {2011})},\ \Eprint
  {http://arxiv.org/abs/1008.0629} {arXiv:1008.0629 [cond-mat.supr-con]}
  \BibitemShut {NoStop}%
\bibitem [{\citenamefont {Gangadharaiah}\ \emph {et~al.}(2011)\citenamefont
  {Gangadharaiah}, \citenamefont {Braunecker}, \citenamefont {Simon},\ and\
  \citenamefont {Loss}}]{MajoranaInteractions}%
  \BibitemOpen
  \bibfield  {author} {\bibinfo {author} {\bibfnamefont {S.}~\bibnamefont
  {Gangadharaiah}}, \bibinfo {author} {\bibfnamefont {B.}~\bibnamefont
  {Braunecker}}, \bibinfo {author} {\bibfnamefont {P.}~\bibnamefont {Simon}}, \
  and\ \bibinfo {author} {\bibfnamefont {D.}~\bibnamefont {Loss}},\ }\href
  {\doibase 10.1103/PhysRevLett.107.036801} {\bibfield  {journal} {\bibinfo
  {journal} {Phys. Rev. Lett.}\ }\textbf {\bibinfo {volume} {107}},\ \bibinfo
  {pages} {036801} (\bibinfo {year} {2011})}\BibitemShut {NoStop}%
\bibitem [{\citenamefont {Stoudenmire}\ \emph {et~al.}(2011)\citenamefont
  {Stoudenmire}, \citenamefont {Alicea}, \citenamefont {Starykh},\ and\
  \citenamefont {Fisher}}]{Miles}%
  \BibitemOpen
  \bibfield  {author} {\bibinfo {author} {\bibfnamefont {E.~M.}\ \bibnamefont
  {Stoudenmire}}, \bibinfo {author} {\bibfnamefont {J.}~\bibnamefont {Alicea}},
  \bibinfo {author} {\bibfnamefont {O.~A.}\ \bibnamefont {Starykh}}, \ and\
  \bibinfo {author} {\bibfnamefont {M.~P. A.}\ \bibnamefont {Fisher}},\ }\href
  {\doibase 10.1103/PhysRevB.84.014503} {\bibfield  {journal} {\bibinfo
  {journal} {Phys. Rev. B}\ }\textbf {\bibinfo {volume} {84}},\ \bibinfo
  {pages} {014503} (\bibinfo {year} {2011})}\BibitemShut {NoStop}%
\bibitem [{\citenamefont {Lutchyn}\ and\ \citenamefont
  {Fisher}(2011)}]{lutchyn_fisher}%
  \BibitemOpen
  \bibfield  {author} {\bibinfo {author} {\bibfnamefont {R.~M.}\ \bibnamefont
  {Lutchyn}}\ and\ \bibinfo {author} {\bibfnamefont {M.~P.~A.}\ \bibnamefont
  {Fisher}},\ }\href {\doibase 10.1103/PhysRevB.84.214528} {\bibfield
  {journal} {\bibinfo  {journal} {Phys. Rev. B}\ }\textbf {\bibinfo {volume}
  {84}},\ \bibinfo {pages} {214528} (\bibinfo {year} {2011})}\BibitemShut
  {NoStop}%
\bibitem [{\citenamefont {Sela}\ \emph {et~al.}(2011)\citenamefont {Sela},
  \citenamefont {Altland},\ and\ \citenamefont {Rosch}}]{Sela}%
  \BibitemOpen
  \bibfield  {author} {\bibinfo {author} {\bibfnamefont {E.}~\bibnamefont
  {Sela}}, \bibinfo {author} {\bibfnamefont {A.}~\bibnamefont {Altland}}, \
  and\ \bibinfo {author} {\bibfnamefont {A.}~\bibnamefont {Rosch}},\ }\href
  {\doibase 10.1103/PhysRevB.84.085114} {\bibfield  {journal} {\bibinfo
  {journal} {Phys. Rev. B}\ }\textbf {\bibinfo {volume} {84}},\ \bibinfo
  {pages} {085114} (\bibinfo {year} {2011})}\BibitemShut {NoStop}%
\bibitem [{\citenamefont {Fidkowski}\ \emph {et~al.}(2011)\citenamefont
  {Fidkowski}, \citenamefont {Lutchyn}, \citenamefont {Nayak},\ and\
  \citenamefont {Fisher}}]{Fidkowski2011}%
  \BibitemOpen
  \bibfield  {author} {\bibinfo {author} {\bibfnamefont {L.}~\bibnamefont
  {Fidkowski}}, \bibinfo {author} {\bibfnamefont {R.~M.}\ \bibnamefont
  {Lutchyn}}, \bibinfo {author} {\bibfnamefont {C.}~\bibnamefont {Nayak}}, \
  and\ \bibinfo {author} {\bibfnamefont {M.~P.~A.}\ \bibnamefont {Fisher}},\
  }\href {\doibase 10.1103/PhysRevB.84.195436} {\bibfield  {journal} {\bibinfo
  {journal} {Phys. Rev. B}\ }\textbf {\bibinfo {volume} {84}},\ \bibinfo
  {pages} {195436} (\bibinfo {year} {2011})}\BibitemShut {NoStop}%
\bibitem [{\citenamefont {Lobos}\ \emph {et~al.}(2012)\citenamefont {Lobos},
  \citenamefont {Lutchyn},\ and\ \citenamefont {Das~Sarma}}]{LobosPRL'12}%
  \BibitemOpen
  \bibfield  {author} {\bibinfo {author} {\bibfnamefont {A.~M.}\ \bibnamefont
  {Lobos}}, \bibinfo {author} {\bibfnamefont {R.~M.}\ \bibnamefont {Lutchyn}},
  \ and\ \bibinfo {author} {\bibfnamefont {S.}~\bibnamefont {Das~Sarma}},\
  }\href {\doibase 10.1103/PhysRevLett.109.146403} {\bibfield  {journal}
  {\bibinfo  {journal} {Phys. Rev. Lett.}\ }\textbf {\bibinfo {volume} {109}},\
  \bibinfo {pages} {146403} (\bibinfo {year} {2012})}\BibitemShut {NoStop}%
\bibitem [{\citenamefont {Kane}\ and\ \citenamefont
  {Fisher}(1992{\natexlab{b}})}]{KaneFisher2}%
  \BibitemOpen
  \bibfield  {author} {\bibinfo {author} {\bibfnamefont {C.~L.}\ \bibnamefont
  {Kane}}\ and\ \bibinfo {author} {\bibfnamefont {M.~P.~A.}\ \bibnamefont
  {Fisher}},\ }\href {\doibase 10.1103/PhysRevB.46.7268} {\bibfield  {journal}
  {\bibinfo  {journal} {Phys. Rev. B}\ }\textbf {\bibinfo {volume} {46}},\
  \bibinfo {pages} {7268} (\bibinfo {year} {1992}{\natexlab{b}})}\BibitemShut
  {NoStop}%
\bibitem [{\citenamefont {Kane}\ and\ \citenamefont
  {Fisher}(1992{\natexlab{c}})}]{KaneFisher3}%
  \BibitemOpen
  \bibfield  {author} {\bibinfo {author} {\bibfnamefont {C.~L.}\ \bibnamefont
  {Kane}}\ and\ \bibinfo {author} {\bibfnamefont {M.~P.~A.}\ \bibnamefont
  {Fisher}},\ }\href {\doibase 10.1103/PhysRevB.46.15233} {\bibfield  {journal}
  {\bibinfo  {journal} {Phys. Rev. B}\ }\textbf {\bibinfo {volume} {46}},\
  \bibinfo {pages} {15233} (\bibinfo {year} {1992}{\natexlab{c}})}\BibitemShut
  {NoStop}%
\bibitem [{\citenamefont {{Kamenev}}\ and\ \citenamefont
  {{Levchenko}}(2009)}]{Kamenev}%
  \BibitemOpen
  \bibfield  {author} {\bibinfo {author} {\bibfnamefont {A.}~\bibnamefont
  {{Kamenev}}}\ and\ \bibinfo {author} {\bibfnamefont {A.}~\bibnamefont
  {{Levchenko}}},\ }\href {\doibase 10.1080/00018730902850504} {\bibfield
  {journal} {\bibinfo  {journal} {Advances in Physics}\ }\textbf {\bibinfo
  {volume} {58}},\ \bibinfo {pages} {197} (\bibinfo {year} {2009})},\ \Eprint
  {http://arxiv.org/abs/0901.3586} {arXiv:0901.3586 [cond-mat.other]}
  \BibitemShut {NoStop}%
\bibitem [{\citenamefont {Affleck}\ and\ \citenamefont
  {Ludwig}(1993)}]{Affleck1993}%
  \BibitemOpen
  \bibfield  {author} {\bibinfo {author} {\bibfnamefont {I.}~\bibnamefont
  {Affleck}}\ and\ \bibinfo {author} {\bibfnamefont {A.~W.~W.}\ \bibnamefont
  {Ludwig}},\ }\href {\doibase 10.1103/PhysRevB.48.7297} {\bibfield  {journal}
  {\bibinfo  {journal} {Phys. Rev. B}\ }\textbf {\bibinfo {volume} {48}},\
  \bibinfo {pages} {7297} (\bibinfo {year} {1993})}\BibitemShut {NoStop}%
\bibitem [{\citenamefont {{Lesage}}\ and\ \citenamefont
  {{Saleur}}(1999)}]{Saleur1999}%
  \BibitemOpen
  \bibfield  {author} {\bibinfo {author} {\bibfnamefont {F.}~\bibnamefont
  {{Lesage}}}\ and\ \bibinfo {author} {\bibfnamefont {H.}~\bibnamefont
  {{Saleur}}},\ }\href {\doibase 10.1016/S0550-3213(99)00076-0} {\bibfield
  {journal} {\bibinfo  {journal} {Nuclear Physics B}\ }\textbf {\bibinfo
  {volume} {546}},\ \bibinfo {pages} {585} (\bibinfo {year} {1999})},\ \Eprint
  {http://arxiv.org/abs/arXiv:cond-mat/9812045} {arXiv:cond-mat/9812045}
  \BibitemShut {NoStop}%
\bibitem [{\citenamefont {Safi}(1997)}]{Safi'97}%
  \BibitemOpen
  \bibfield  {author} {\bibinfo {author} {\bibfnamefont {I.}~\bibnamefont
  {Safi}},\ }\href@noop {} {\bibfield  {journal} {\bibinfo  {journal} {ANNALES
  DE PHYSIQUE}\ }\textbf {\bibinfo {volume} {22}},\ \bibinfo {pages} {463}
  (\bibinfo {year} {1997})}\BibitemShut {NoStop}%
\end{thebibliography}

%
\end{document}